\documentclass[a4paper,11pt]{article}
\pdfoutput=1 

\usepackage{jinstpub} 

\usepackage{lineno}
\usepackage{float}

\title{\boldmath Optical calibration of the SNO+ detector in the water phase with deployed sources}

\collaboration{%
	\includegraphics[height=17mm]{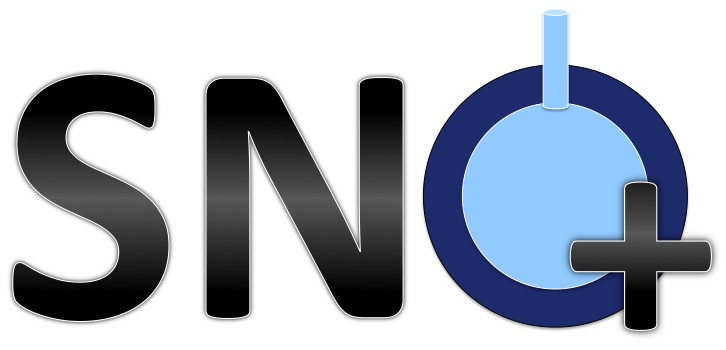}\\[6pt]
	SNO+ collaboration}
\author[a]{M.\,R.\,Anderson,}
\author[b]{S.\,Andringa,}
\author[c,d]{M.\,Askins,}
\author[e]{D.\,J.\,Auty,}

\author[b,f]{F.\,Bar\~{a}o,}
\author[b,g]{N.\,Barros,}
\author[h]{R.\,Bayes,}
\author[i]{E.\,W.\,Beier,}
\author[j,h]{A.\,Bialek,}
\author[k]{S.\,D.\,Biller,}
% ...

\affiliation[a]{\it Queen's University, Department of Physics, Engineering Physics \& Astronomy, Kingston, ON K7L 3N6, Canada}
\affiliation[b]{\it Laborat\'{o}rio de Instrumenta\c{c}\~{a}o e  F\'{\i}sica Experimental de Part\'{\i}culas (LIP), Av. Prof. Gama Pinto, 2, 1649-003, Lisboa, Portugal}
\affiliation[c]{\it University of California, Berkeley, Department of Physics, CA 94720, Berkeley, USA}
\affiliation[d]{\it Lawrence Berkeley National Laboratory, 1 Cyclotron Road, Berkeley, CA 94720-8153, USA}
\affiliation[e]{\it University of Alberta, Department of Physics, 4-181 CCIS,  Edmonton, AB T6G 2E1, Canada}
\affiliation[f]{\it Universidade de Lisboa, Instituto Superior T\'{e}cnico (IST), Departamento de F\'{\i}sica, Av. Rovisco Pais, 1049-001 Lisboa, Portugal}
\affiliation[g]{\it Universidade de Lisboa, Faculdade de Ci\^{e}ncias (FCUL), Departamento de F\'{\i}sica, Campo Grande, Edif\'{\i}cio C8, 1749-016 Lisboa, Portugal}
\affiliation[h]{\it Laurentian University, Department of Physics, 935 Ramsey Lake Road, Sudbury, ON P3E 2C6, Canada}
\affiliation[i]{\it University of Pennsylvania, Department of Physics \& Astronomy, 209 South 33rd Street, Philadelphia, PA 19104-6396, USA}
\affiliation[j]{\it SNOLAB, Creighton Mine \#9, 1039 Regional Road 24, Sudbury, ON P3Y 1N2, Canada}
\affiliation[k]{\it University of Oxford, The Denys Wilkinson Building, Keble Road, Oxford, OX1 3RH, UK}
% ...

\emailAdd{ainacio@lip.pt}

\abstract{SNO+ is a large-scale liquid scintillator experiment with the primary goal of searching for neutrinoless double beta decay, and is located approximately 2 km underground in SNOLAB, Sudbury, Canada. The detector acquired data for two years as a pure water Cherenkov detector, starting in May 2017. During this period, the optical properties of the detector were measured \textit{in situ} using a deployed light diffusing sphere, with the goal of improving the detector model and the energy response systematic uncertainties. The measured parameters included the water attenuation coefficients, effective attenuation coefficients for the acrylic vessel, and the angular response of the photomultiplier tubes and their surrounding light concentrators, all across different wavelengths. The calibrated detector model was validated using a deployed tagged gamma source, which showed a 0.6\% variation in energy scale across the primary target volume.}


\arxivnumber{2106.03951} 

\begin{document}
\maketitle
\flushbottom

\section{Introduction}
\label{section:introduction}

SNO+ is a multi-purpose neutrino experiment with the primary goal of searching for neutrinoless double beta decay of $^{130}$Te \cite{snoplus-prospects}. The detector re-uses most of the components of the Sudbury Neutrino Observatory (SNO) that operated from 1999 to 2006 \cite{sno-detector}, with several major upgrades to enable the use of liquid scintillator as target material. It consists of a spherical acrylic vessel (AV) with a thickness of 55 mm and a radius of 6 m, surrounded by a geodesic steel structure that holds 9362 inward-facing photomultiplier tubes (PMTs) at an average distance of 8.35 m from the center of the AV. The PMTs are equipped with light concentrators, yielding an effective optical coverage of approximately 54\%. A 6.8-m tall acrylic cylinder of 0.75 m radius extends from the top of the AV, providing access for the deployment of calibration sources. The volume outside the AV, including the 22-m wide and 34-m high cavity into which the detector is inserted, is filled with 7000 tonnes of ultra-pure water that shields against the radioactivity from the instrumentation and surrounding rock. A full description of the detector can be found in \cite{sno-detector,snoplus-detector}.

SNO+ acquired data as a pure-water Cherenkov detector between May 2017 and June 2019. The Water Phase served as a commissioning stage for upgraded readout electronics prior to the filling of the detector with liquid scintillator loaded with Tellurium. In all stages of the experiment, the main observables of SNO+ are the times at which PMTs first detect a photon and the charge collected within a time window, from which estimators of energy, position, direction, and particle ID are reconstructed.
Since photons produced inside the detector must propagate through multiple optical media to reach the PMTs, and their collection is affected by the optical properties of the PMT and light concentrators, the parameters for light propagation in the detector must be carefully understood and monitored to yield a precise reconstruction of the events occurring throughout the detector volume. This is accomplished by ensuring the detector media are clean and transparent, and by measuring \textit{in situ} the optical properties of the PMTs and concentrators, the target medium (ultra-pure water or scintillator), the acrylic, and the external water surrounding the AV.

The optical calibration during the water phase was fundamental to establish our knowledge of the optical properties of the detector and evaluating how the concentrators around the PMTs have changed since the transition from SNO to SNO+. Having water both inside and outside the AV also provided a unique opportunity to accurately measure the properties of the external water and acrylic, before the transition to using scintillator as the target medium. These optical properties were measured across a range of wavelengths using a light diffusing sphere ("laserball"), deployed in several positions inside and outside the AV. The measurements were used to calibrate the detector simulation model, which was then validated using a gamma source.

This paper discusses the optical calibration of the SNO+ detector in the water phase. Section \ref{section:sources} briefly presents the calibration sources used, Section \ref{section:method} describes the optical calibration analysis method, Sections \ref{section:data}--\ref{section:results} describe the water phase calibration campaigns and report the measurements performed. Finally, Section \ref{section:n16validation} describes the validation of the measurements using a gamma source and the implications for the SNO+ energy scale uncertainty.


\section{Motivation and goal of the optical calibration}
\label{section:motivation}

In all phases of the SNO+ experiment, both physics events of interest and undesired background events within the detector will create light that will propagate through the detector. As it propagates, the light will be subject to optical processes such as refraction, reflection, absorption, and a variety of wavelength-dependent scattering interactions. These effects are governed by the properties of the materials in the detector: the water or scintillator inside the AV, the acrylic of the AV itself and the water outside the AV. Additionally, the sensitivity of the detector to light from different positions throughout its volume depends on the combined efficiency of the PMTs and their surrounding light concentrators as a function of wavelengths and incident angle.

The energies of the events are determined by the spatial and temporal distribution of the PMTs that detect photons ("hit" PMTs), as well as the total number of hits. Due to the aforementioned optical effects, an event near the inner surface of the AV produces a significantly different number of hits than a similar event near the center of the AV. An \textit{in situ} measurement of the optical properties of the SNO+ detector is essential for a realistic model describing the propagation and detection of light from all sources and to minimize the uncertainty of the absolute energy scale and the reconstructed positions.

\begin{figure}[htbp]
	\centering
	\includegraphics[width=.6\textwidth]{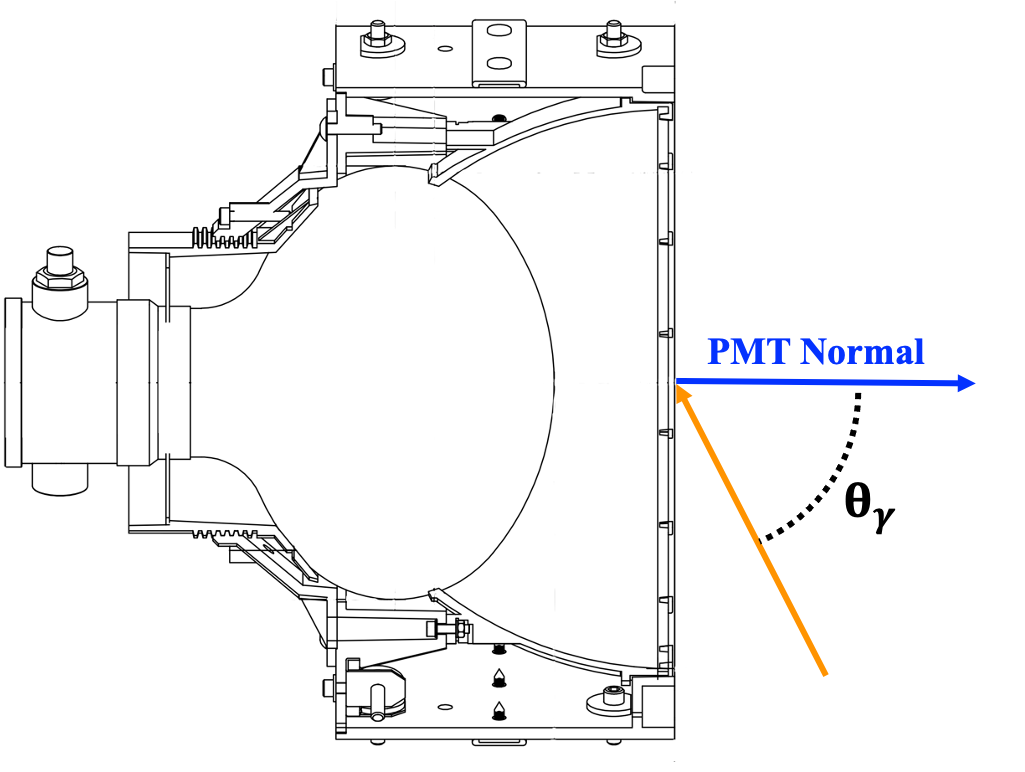}
	\caption{\label{fig:pmtscheme} Technical diagram of the PMT and concentrator assembly. The incident angle $\theta_\gamma$ is the angle that the incident light (orange) makes with the assembly entrance, defined by a normal vector (blue).}
\end{figure}

The measured optical parameters are inputs for the Monte Carlo detector model, thoroughly described in \cite{sno-d2o}. The optics of the detector media is modeled by arrays of absorption and scattering lengths as a function of wavelength. The detector simulation has two available models for the PMTs and their associated light concentrators (Figure \ref{fig:pmtscheme}): a detailed three dimensional model \cite{sno-d2o}, and a simplified empirical model called the grey disc model. While the former models all the interactions of light with the PMT and concentrator geometry, the latter replaces the complex geometry with a flat disc at the front opening of the concentrator support structure. When a photon reaches the disc, instead of modeling all its interactions in the structure, the grey disc model assigns a reflection and absorption probability to the contact point, based on the incident angle and wavelength. The grey disc is the preferred PMT model in SNO+ because it speeds up the time of the simulation and its optical properties are calibrated directly from the optical calibration measurements.

Attenuation lengths are measured with the laserball, and scattering lengths are measured with a fixed system of optical fibers described in \cite{snoplus-detector}. These combine to estimate a set of absorption lengths for a range of wavelengths. The laserball measures the angular response of the PMTs and concentrators, i.e. the combined efficiency in collecting light with a given incident angle, $\theta_\gamma$, relative to normal incidence, $\theta_\gamma = 0$. These measurements are directly converted into grey disc model absorption probabilities. Both models depend on a scaling factor for the PMT collection efficiency, which is the probability for a generated photo-electron to successfully reach the first dynode, resulting in a signal that is propagated to the front-end electronics. This factor is tuned by comparing simulations to data of calibration sources deployed at the center of the detector. Calibration data are also used to adjust reflectivity parameters of the PMT models, necessary to correctly reproduce the time distribution of light arriving to the PMTs at times later than direct photons.

The energy of an event in water is estimated from the amount of detected direct light, i.e. the number of hit PMTs within a time interval between -10 and 8 ns centered on the event time after correcting for the time-of-flight.\footnote{The energy reconstruction uses a time window wider than the optical calibration time window, described in Section \ref{section:method}, to maximize the number of hits available for reconstruction, without needing to include significant corrections for scattered or reflected photons.} This approach is used in order to avoid the late light region, which is harder to model with the same accuracy because of the reflections from the detector components and from the PMTs and their concentrators. The absolute energy scale is determined by mono-energetic calibration sources (like the $^{16}$N tagged gamma source) at the center of the detector, where the detector properties are most symmetric. The primary estimate of the systematic uncertainty on the energy scale is the volume-weighted average difference between the Monte Carlo model prediction of the detector response to the calibration source and the source data itself. With a thorough calibration of the detector, SNO+ aims to minimize the energy scale systematic to $\leq$ 1\%, as was the case for the SNO detector \cite{sno-d2o,sno-salt}. Additionally, the tuned Monte Carlo model is expected to correctly reproduce the arrival time distributions of the photons, which is necessary to create accurate probability density functions for the position reconstruction algorithms.


\section{Deployed calibration sources} 
\label{section:sources}

During the SNO+ water phase, the laserball was used to determine the optical properties of the detector, necessary to correctly account for the detector's energy response variation with position, and the $^{16}$N tagged gamma source was used to determine the energy scale.
In a calibration campaign, data are collected with a calibration source placed in a specific position inside the detector over the course of a run (with a specific time length). A set of runs with the source in different positions is known as a calibration scan. A typical laserball optical calibration scan include $\sim$40 internal positions. Using a manipulator system, calibration sources can be deployed in many positions within two orthogonal planes inside the acrylic vessel, as well as in the water region between the vessel and the PMTs along a few vertical axes. Section \ref{section:data} provides detailed information about the calibration campaigns which provided the data analyzed in this paper. The manipulator system and other calibration hardware are discussed in detail in \cite{snoplus-detector}.

\subsection{Laserball}
\label{subsection:laserball}

\begin{figure}[htbp]
	\centering
	\includegraphics[width=0.8\textwidth]{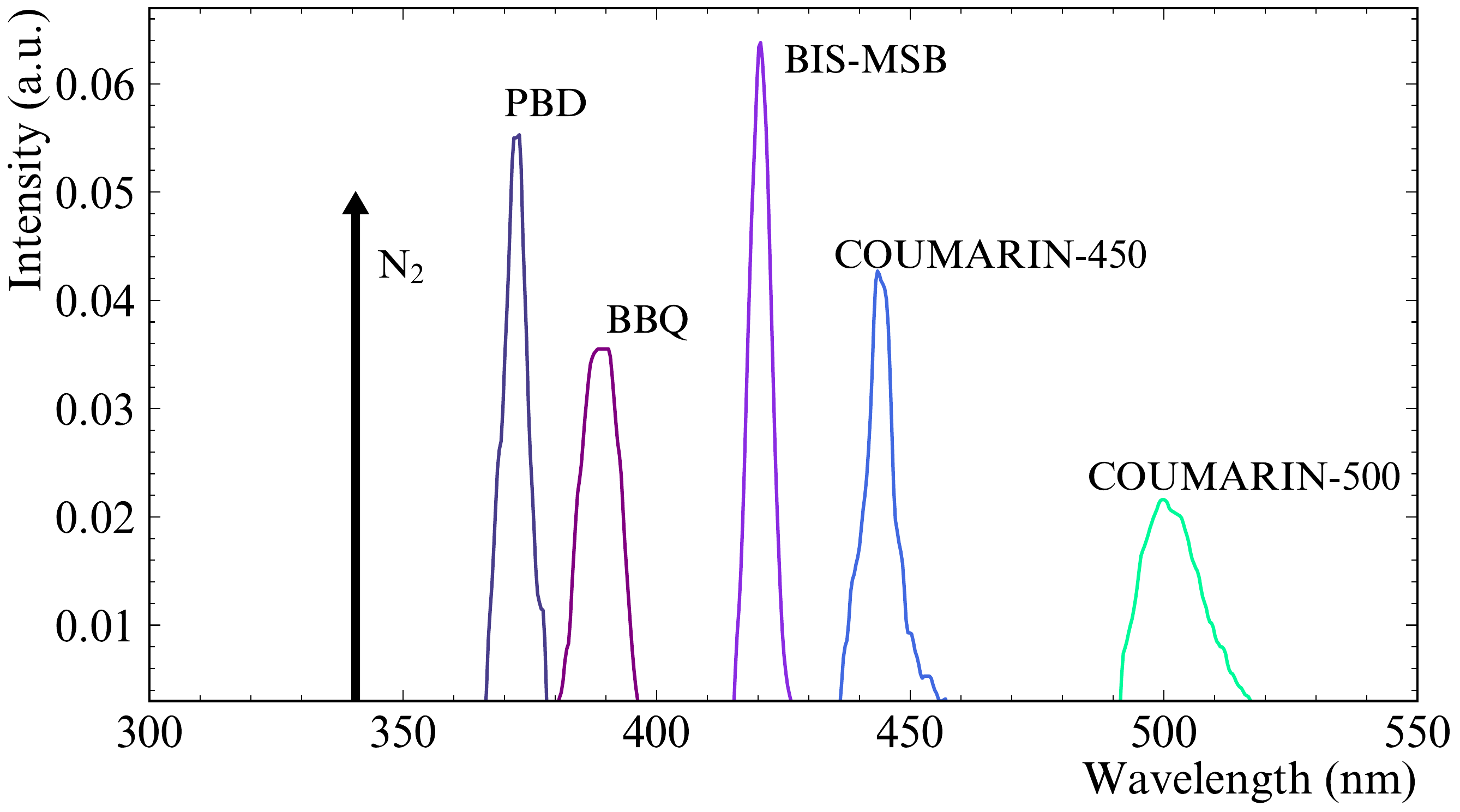}
	\caption{\label{fig:lbDyes} Wavelength spectra of the N$_2$ laser and of the dyes used during the SNO+ water phase.}
\end{figure}

The main optical calibration source used during the SNO+ water phase was a light diffusing sphere, the laserball, inherited from SNO \cite{OpCalHardware}. It consists of an $\sim$11 cm diameter spherical quartz flask filled with small air-filled glass beads (50 $\mu$m in diameter) suspended in silicone gel. The beads diffuse light injected into the flask through a fiber guide. The light comes from a nitrogen pumped dye laser system, located in the deck clean room (DCR) above the detector. In addition to the primary wavelength of the laser (337.1 nm), five selected dyes provide additional wavelength ranges centered at 365 (PBD), 385 (BBQ), 420 (BIS-MSB), 450 (Coumarin-450) and 500 nm (Coumarin-500). Figure \ref{fig:lbDyes} shows the stimulated emission spectra of each of the dyes, measured directly from the calibration laser system with an Ocean Optics USB 2000+ UV-VIS Spectrometer.

A photodiode close to the laser produces a timing signal that triggers the data acquisition system. The laser beam intensity is controlled by the use of two successive sets of neutral density filters mounted in rotating supports. The light then passes through about 30 m of optical fiber to the laserball.
Although the laserball was designed to be an isotropic light source, the mounting hardware on top of the flask partially shadows the light going upwards, reducing the intensity of the light traveling in this direction by about 50\%. As will be discussed in Section \ref{section:method}, the overall anisotropy of the laserball is important to consider when interpreting laserball data, and it is part of the information extracted from the optical calibration analysis.

\begin{figure}[htbp]
	\centering
	\includegraphics[width=1\textwidth]{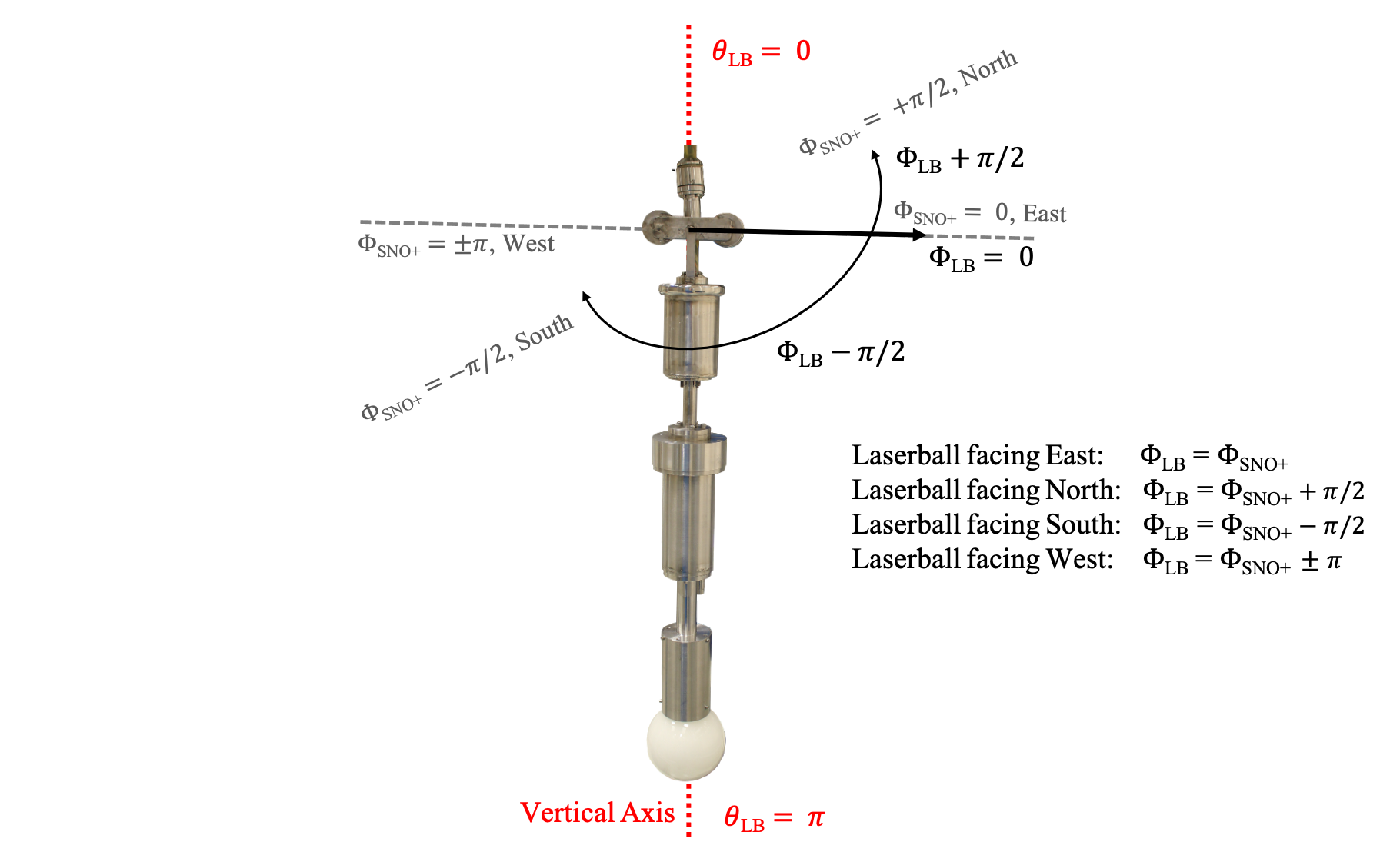}
	\caption{\label{fig:lbcoordinates} Schematic of the laserball coordinate system $(\text{cos}\theta_{\text{LB}},\Phi_{\text{LB}})$, defined by the direction of a reference mark in the laserball hardware, represented by the black arrow. When the laserball rotates, its coordinate system changes relative to the fixed SNO+ detector coordinate system $(\text{cos}\theta_{\text{SNO+}},\Phi_{\text{SNO+}})$. Since the laserball does not rotate along $\theta$, $\text{cos}\theta_{\text{SNO+}}$ is equivalent to $\text{cos}\theta_{\text{LB}}$. The laserball coordinate system coincides with the detector coordinate system when $\Phi_{\text{LB}}$ points towards East.}
\end{figure}

When deployed internally, the laserball is attached to a pair of side ropes which physically constrain it to have one of four possible azimuthal orientations (north, south, east and west). The orientations are relative to the fixed detector coordinate system, and the relations between the two coordinate systems are illustrated in Figure \ref{fig:lbcoordinates}. When outside of the AV, the laserball orientation is not constrained and has to be determined afterwards, as will be discussed in Section \ref{subsec:preFit}.

\subsection{The $^{16}$N tagged gamma source}
\label{subsection:n16}

The main energy calibration source used during the SNO+ water phase was the $^{16}$N gamma-ray source, also inherited from SNO \cite{sno-n16,sno-detector}. This calibration source was used to determine the energy scale and the reconstruction systematics. $^{16}$N nuclei (t$_{1/2}$ = 7.13 s, Q-value = 10.42 MeV) are produced in a shielded pit near the detector cavity by bombarding $^{16}$O, in gaseous CO$_2$, with 14 MeV neutrons from a Deuterium-Tritium (DT) generator. The activated gas is then transported into a decay chamber deployed in the SNO+ water volume.
There, the $^{16}$N beta-decays to $^{16}$O$^*$ (B.R. 66.2\%), which de-excites emitting a 6.1 MeV gamma. There are two other decay branches: one that produces 7.1 MeV gamma-rays in coincidence with the beta (6\%), and a direct branch to the ground state (28\%), resulting in a 10.4 MeV endpoint beta-particle. 

The decay chamber was designed to contain the energetic beta-particles. They interact with plastic scintillator lining the walls of the chamber volume, creating optical scintillation photons that are measured with the SNO+ electronics and provide a tag to select $^{16}$N events. The gamma-rays are able to exit the chamber and interact via Compton scattering in the detector medium to produce high energy electrons, which in turn produce Cherenkov photons that are observed and result in a broad (3--7 MeV) reconstructed spectrum. 


\section{Optical calibration analysis method}
\label{section:method}

\begin{figure}[htbp]
	\centering
	\includegraphics[width=.31\textwidth]{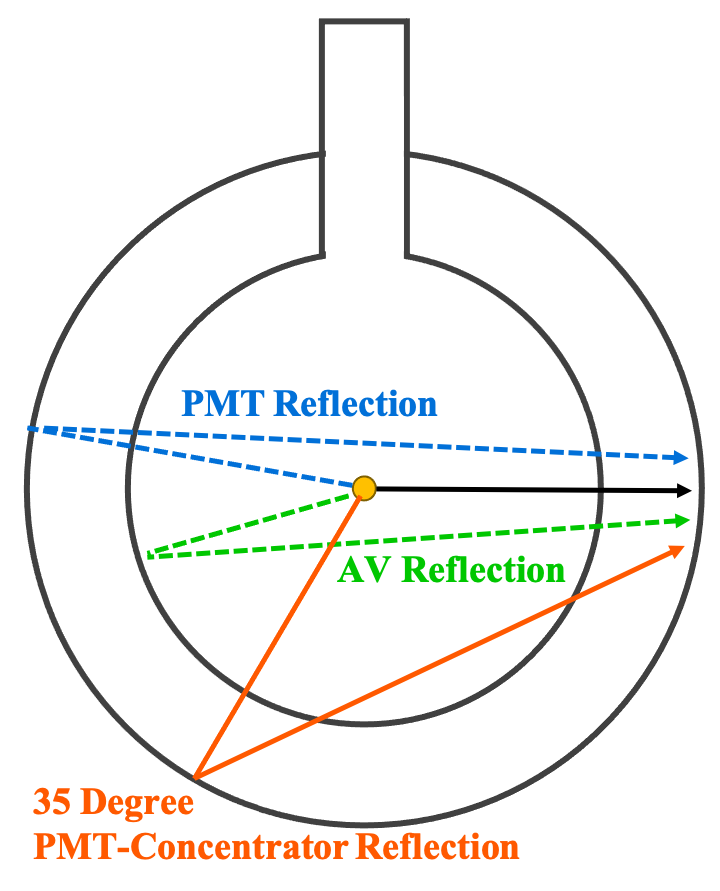}
	\includegraphics[width=.68\textwidth]{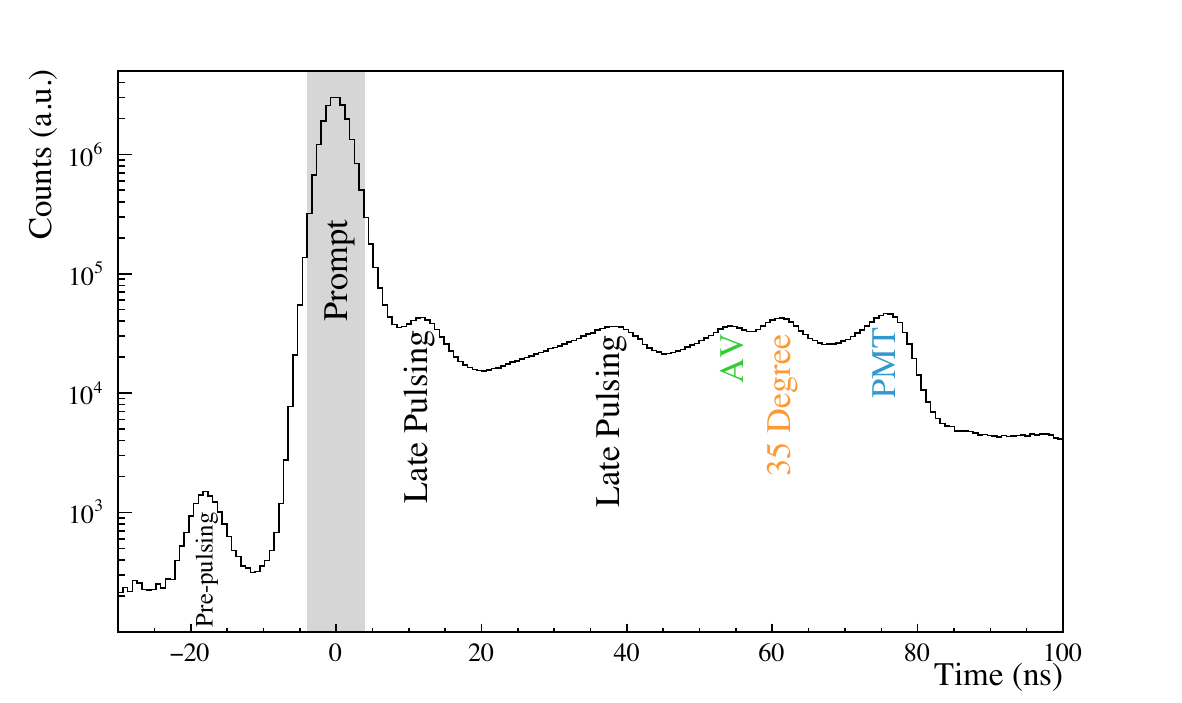}
	\caption{\label{fig:lightPaths_timeResid} Left: Optical paths within the detector for a central laserball position. The black line represents direct light; the blue line represents light reflected by the PMT glass bulb and multiply-reflected by the PMT concentrators; the green line represents light reflected by the AV boundary; and the orange line represents light reflected off of the concentrators surrounding the PMTs. Right: PMT time distribution for a central laserball data run. The shaded region corresponds to the $\pm$4 ns prompt time residual window used for the optical calibration analysis. This approach is used in order to avoid the late light region, which is harder to model with the same accuracy. The pre- and late-pulsing are features of the PMT time response, as identified in \cite{sno-pmt}.}
\end{figure}

The optical calibration analysis of the laserball data only considers light arriving from the source to the PMTs in a narrow $\pm$4 ns time residual window, centered around the prompt peak, shown in Figure \ref{fig:lightPaths_timeResid}. The time residual, $t_{\textrm{res}}$, is the instantaneous event time which accounts for the light propagation time to the PMT, $t_{\textrm{TOF}}$, relative to the PMT hit time, $t_{\textrm{PMT}}$, and a constant time offset, $t_0$:
\begin{equation}
\label{eq:timeresidual}
t_{\textrm{res}} = t_{\textrm{PMT}}-t_0-t_{\textrm{TOF}}\:.
\end{equation}
Using the prompt light allows the accurate characterization of the optical parameters without requiring detailed knowledge of the geometry and reflective properties of the PMTs, concentrators and other components in the detector, which strongly impact late light. 

The PMTs register only single hits even when multiple photoelectrons (MPE) are produced in the PMT from a single laser pulse. Using simply the prompt hit count for each PMT \emph{j}, $N_{ij}$, would then underestimate the photon intensity of the laserball during a run \emph{i}. To take into account the probability of MPE hits, the number of
prompt counts $N_{ij}$ is corrected by inverting the expected
Poisson distribution of the hit counts with mean $\xi_{ij}$:
\begin{equation}
\begin{split}
\text{Prob}\:(1 \:\text{hit}) &= \text{Prob}\:(\geq 1 \:\text{photoelectron}) = 1 - \text{Prob}\:(0 \:\text{photoelectrons}) = \frac{N_{ij}}{N^{pulses}_i}\\
&\Longrightarrow \frac{N_{ij}}{N^{pulses}_i}=1-\frac{(\xi_{ij})^0e^{-\xi_{ij}}}{0!}=1-e^{-\xi_{ij}}\\
&\Longrightarrow \xi_{ij} = - \ln\left(1-\frac{N_{ij}}{N^{pulses}_i}\right) = \frac{N^{\textrm{MPE}}_{ij}}{N^{pulses}_i}\:, \\
\end{split}
\label{eq:meanPoisson}
\end{equation}
where $N^{\textrm{MPE}}_{ij}$ is proportional to the actual number of prompt photons that strike the PMT (valid for small numbers of incident photons), and $N^{pulses}_i$ is the number of laser pulses during a laserball run \textit{i}. The optical calibration analysis uses the occupancy $O^{\textrm{data}}_{ij}$ measured by PMT \emph{j} during a run \emph{i}, with the laserball at a given position emitting light at a single wavelength, which is:
\begin{equation}
O^{\textrm{data}}_{ij} = \frac{N^{\textrm{MPE}}_{ij}}{N_i^{pulses}}\:.
\end{equation}

The measured $O^{\textrm{data}}_{ij}$ relates to the optical properties of the detector through a model based on geometrical optics --- it assumes that the detector can be characterized by averaging some properties, such as considering that the media is homogeneous and isotropic and that the PMT-concentrator assembly response depends only on the incident angle of light. 
The model parameterizes the expected occupancy observed by PMT \emph{j} during a run \emph{i}, $O^{\textrm{model}}_{ij}$, as follows \cite{sno-d2o}:
\begin{equation}
\label{eq:optmodel}
\centering	
O^{\textrm{model}}_{ij}=N_i\Omega_{ij}R_{ij}T_{ij}L_{ij}\epsilon_{j}e^{-\left(d^{\text{w}_{\text{int}}}_{ij}\alpha_{\text{w}_{\text{int}}}+d^\text{a}_{ij}\alpha_\text{a}+d^{\text{w}_{\text{ext}}}_{ij}\alpha_{\text{w}_{\text{ext}}} \right)}\:,
\end{equation}
where the terms are defined as:
\begin{itemize}
	\setlength{\itemsep}{1pt}
	\item $N_{i}$ --- number of photons emitted by the laserball in run \emph{i}, and detected within a prompt timing window by all PMTs. This term is the intensity normalization for the run;
	\item $\Omega_{ij}$ --- solid angle subtended by the PMT-concentrator assembly \emph{j} from the laserball position in run \emph{i};
	\item $R_{ij}$ --- PMT and concentrator angular response beyond the solid angle $\Omega_{ij}$. This factor is parameterized as a function of the photon incident angle on the face of the PMT-concentrator assembly;
	\item $T_{ij}$ --- Fresnel transmission coefficients for the media interfaces, calculated from the refractive indices, wavelengths and incidence angles of light at the boundaries;
	\item $L_{ij}$ --- the laserball light intensity distribution, parameterized as a function of the polar ($\textrm{cos}\theta_{\text{LB}}$) and azimuthal ($\mathit{\phi_{\text{LB}}}$) angles of the light ray relative to the laserball center. This parameter is included in the model to account for the small anisotropies in the laserball light emission;
	\item $\epsilon_{j}$ --- relative efficiency of PMT \emph{j}, combining the overall PMT efficiency and electronics threshold effects (including the quantum efficiency (QE), which refers to the wavelength-dependent probability of registering a hit);
	\item $d^{\text{w}_{\text{int}},\text{a},\text{w}_{\text{ext}}}_{ij}$ --- refracted light path lengths through the internal water ($d^{\text{w}_{\text{int}}}$), the acrylic ($d^\text{a}$) and external water ($d^{\text{w}_{\text{ext}}}$);
	\item $\alpha_{\text{w}_{\text{int}},\text{a},\text{w}_{\text{ext}}}$ --- attenuation coefficients for the internal water ($\alpha_{\text{w}_{\text{int}}}$), the acrylic ($\alpha_\text{a}$) and external water ($\alpha_{\text{w}_{\text{ext}}}$).
\end{itemize}

 The solid angles, $\Omega_{ij}$, the Fresnel transmission coefficients, $T_{ij}$, and the refracted light path lengths in each medium, $d_{ij}$, are determined simply from the laserball and PMT positions, and the detector geometry. The remaining parameters are extracted from the laserball data through a multi-parameter fit described in Section \ref{subsec:ocaFit}.

By building a data set which includes many different laserball positions, it is possible to break covariances between the model parameters, such as between $\alpha_w$ and $R_{ij}$. However, the distances through the external water and the acrylic are correlated for laserball positions inside the AV. Therefore, for data taken only inside the AV, the covariance between $\alpha_w$ and $\alpha_a$ is difficult to break and, typically, previous measurements for acrylic attenuation are used as fixed inputs to the optical model and only $\alpha_w$ is fit. 

When adding data from laserball positions outside the AV, it becomes easier to disentangle the correlation between the acrylic and the external water, allowing both parameters to be extracted simultaneously. External positions also probe higher incidence angles at the PMTs to characterize $R_{ij}$ over a wider range of angles, which is useful to improve the models of the PMTs and the concentrators. However, when the laserball is very close to the PMTs and the AV boundary, there are optical paths that make it very difficult to separate light reflected off of the AV and PMTs from the direct light. For this reason, in the water phase analysis of laserball positions outside the AV, only PMTs whose light paths are fully contained in the external water volume, and within a given angular aperture from the laserball, were considered, as will be discussed in more detail in Section \ref{section:dataSeletion}.

\subsection{Determining the laserball position, light distribution, and orientation}
\label{subsec:preFit}

Many physical quantities in the optical model of Equation \ref{eq:optmodel} depend directly on the accurate determination of the laserball position. Although the source positioning system can provide an estimate of the laserball position, its positioning algorithm is based on the tension and length of the ropes that support the source, which have large uncertainties for positions away from the center of the AV. The laserball position used in this analysis is extracted from the data through a $\chi^2$ minimization of the time residual in Equation \ref{eq:timeresidual}, using the mean of the hit times and its uncertainty for each PMT in a given run.

The laserball is a 4$\pi$ quasi-uniform light source and the anisotropies of its intensity distribution depend on the density of glass beads in the silicone gel, on the positioning of the quartz rod inside the source, and shadowing from the metal parts. Knowing those details is not needed, as we characterize those anisotropies with in-situ laserball data taken at the center of the AV. The intensity distribution is described by twelve independent sinusoidal distributions $H$ in the azimuthal angle $\mathit{\phi_{\text{LB}}}$, each valid in a given range in the polar angle $\theta_{\text{LB}}$ with different amplitudes and phases. The azimuthal distributions are weighted by a single polynomial function in $\textrm{cos}\theta_{\text{LB}}$: 
\begin{equation}
L(\textrm{cos}\theta_{\text{LB}},\mathit{\phi}_{\text{LB}}) = H(\textrm{cos}\theta_{\text{LB}},\mathit{\phi}_{\text{LB}}) \times P(\textrm{cos}\theta_{\text{LB}})\:.
\end{equation}
This particular choice of functions was empirically guided by prior studies in SNO \cite{barros,olivier}. The parameters of the sinusoidal distribution (amplitudes and phases) are measured by analyzing the data from rotated laserball runs (relative to the fixed SNO+ coordinate system) taken at the center of the AV. For a $\textrm{cos}\theta_{\text{LB}}$ slice (equivalent to $\textrm{cos}\theta_{\text{SNO+}}$, since the laserball always has the same vertical orientation), the occupancy ratio of the PMTs in runs with opposite orientations (180$^{\circ}$ apart) is given by:
\begin{equation}
\frac{O_{1j}}{O_{2j}}=\frac{N_1\:\Omega_{1j}\:R_{1j}\:T_{1j}\:L_{1j}\:\epsilon_j\:e^{-(d^{w_{\text{int}}}_{1j}\alpha_{w_{\text{int}}}+d^a_{1j}\alpha_a+d^{w_{\text{ext}}}_{1j}\alpha_{w_{\text{ext}}})}}{N_2\:\Omega_{2j}\:R_{2j}\:T_{2j}\:L_{2j}\:\epsilon_j\:e^{-(d^{w_{\text{int}}}_{2j}\alpha_{w_{\text{int}}}+d^a_{2j}\alpha_a+d^{w_{\text{ext}}}_{2j}\alpha_{w_{\text{ext}}})}}\:.
\end{equation}
All the terms, except the normalizations $N$ and the intensity distribution $L$, are the same for the two runs since the source is in the same position. Hence, the ratio becomes:
\begin{equation}
\frac{O_{1j}}{O_{2j}} = \frac{N_1\times P(\textrm{cos}\theta_{\text{LB}})\times H(\mathit{\phi}_{\text{LB}}+\Phi_1)}{N_2\times P(\textrm{cos}\theta_{\text{LB}})\times H(\mathit{\phi}_{\text{LB}}+\Phi_2)}\:,
\end{equation}
\noindent where $\Phi_1$ and $\Phi_2$ are the relative orientations of the laserball in the two runs, and $L(\textrm{cos}\theta_{\text{LB}},\mathit{\phi}_{\text{LB}})$ is expanded into the sinusoidal function $H$ and the independent polar variation $P$. The latter is the same in the numerator and denominator, thus enabling sensitivity to the azimuthal sinusoidal distribution. The ratios are fitted for all the $\textrm{cos}\theta_{\text{LB}}$, and the extracted sinusoidal parameters are used as the seed to the main optical calibration analysis fit.

Additionally, an independent analysis was developed to extract the laserball orientation in each external position, necessary to correctly describe its light intensity distribution, $L_{ij}$. For external runs, without side ropes attached, an LED was installed to determine the orientation of the laserball. From the coordinates of the region of PMTs with maximum integrated number of hits in the LED runs and the laserball position, it was possible to determine the direction of the LED relative to the detector reference axes. The LED was mounted at a known angle from the laserball reference axes, and by knowing its direction, it was then possible to determine the laserball orientation relative to the detector. The orientations obtained at the different external positions are used as input to the optical calibration analysis fit. This analysis is able to determine the orientation with a precision of $\sim$10 degrees.\footnote{The precision of the laserball orientation is obtained from the difference between the orientations determined using the LED runs before and after data taking at each position.} This precision is sufficient for the optical calibration analysis fit since the laserball intensity asymmetry with $\mathit{\phi}_{\text{LB}}$ is at most 3\%, and an uncertainty of 10 degrees in the source orientation would only affect the PMT occupancy by less than 0.1\%. Typically, LED runs are taken before and after taking laserball data at each position. Differences between the orientations obtained from the LED runs taken before and after data taking indicate whether the laserball rotated.

\subsection{Optical calibration analysis fit}
\label{subsec:ocaFit}

In order to extract the optical parameters in Equation \ref{eq:optmodel} from the laserball data, we use a method that normalizes the occupancy at a PMT \emph{j} for a given run \emph{i}, $O_{ij}$, by the value from a run with the laserball located at the center of the detector, $O_{0j}$ \cite{sno-d2o}. This normalization is done for both the model and for the data as shown in Equations \ref{eq:modelratio} and \ref{eq:dataratio}.
\begin{equation}
\label{eq:modelratio}
Q_{ij}^{\textrm{model}}=\frac{O_{ij}^{\textrm{model}}}{O_{0j}^{\textrm{model}}}\left(\frac{\Omega_{0j}T_{0j}}{\Omega_{ij}T_{ij}}\right)=\frac{N_iR_{ij}L_{ij}}{N_0R_{0j}L_{0j}}exp\left(-\sum_{k}\left(d_{ij}^k-d_{0j}^k\right)\alpha_k\right)\:.
\end{equation}
\begin{equation}
Q_{ij}^{\textrm{data}}=\frac{O_{ij}^{\textrm{data}}}{O_{0j}^{\textrm{data}}}\left(\frac{\Omega_{0j}T_{0j}}{\Omega_{ij}T_{ij}}\right)\:.
\label{eq:dataratio}
\end{equation}
The ratios $Q_{ij}$ are occupancy ratios corrected by the solid angles $\Omega_{ij}$, $\Omega_{0j}$, and Fresnel transmission coefficients $T_{ij}$, $T_{0j}$, which are numerically calculated a priori.
By taking the ratio between an off-center and a central laserball run, the dependency on the PMT efficiency, $\epsilon_{j}$, is removed, eliminating one parameter for each PMT (about 9000) from the model. 

With the exception of the distances $d_{ij}^k$, the model occupancy ratio is entirely characterized by parameters that can be determined by the minimization of a $\chi^2$ estimator over several iterations \cite{sno-d2o}:
\begin{equation}
\chi^2 = \sum_{i}^{\#Runs}\sum_{j}^{\textrm{\#PMTs}}\frac{(Q^{\textrm{data}}_{ij}-Q^{\textrm{model}}_{ij})^2}{\sigma^2_{\textrm{stat},ij}+\sigma^2_{\textrm{PMT}}(\theta_{\gamma,ij})}\:,
\label{eq:occRatioChiSquare}
\end{equation}
\noindent where $\sigma^2_{\textrm{stat},ij}$ is the statistical uncertainty on the data occupancy ratio due to counting statistics, and $\sigma^2_{\textrm{PMT}}(\theta_{\gamma,ij})$ is an additional uncertainty introduced to account for variations in the PMT angular response as a function of the incidence angle of the light. The number of model parameters in the $\chi^2$ is around 166: 3 attenuations, 90 PMT response bins for $R$, 4 coefficients for the laserball $P(\textrm{cos}\theta_{\text{LB}})$ function and 24 parameters for $H(\textrm{cos}\theta_{\text{LB}},\phi_{\text{LB}})$, and 45 run intensity normalizations $N_i$ (average number of laserball data runs of a given wavelength in the fit). Typically, the minimization is performed with more than 100,000 data points, after applying data selection cuts (discussed in Section \ref{section:dataSeletion}), allowing to determine the optical model parameters with a statistical uncertainty below 1\%.

The minimization of the $\chi^2$ is a non-linear least squares problem that is solved using the Levenberg-Marquardt algorithm \cite{levenberg,marquart}. The minimization is performed over several iterations with a sequentially decreasing upper chi-square limit. After each minimization, PMTs with a $\chi^2$ larger than the new limit are removed from the sample (this removes between 10 and 35\% of the data points from each run, depending on the source position). 

The $\chi^2$ cut removes PMTs in which some aspect of the optics is not modeled well; for instance, PMTs undergoing irregular exposure to light due to scattering or reflections which are unaccounted for by the model. To avoid a sequential minimization over the same subset of the sample, all PMTs, even those previously removed, are reconsidered in each iteration. The minimization is performed using all the laserball data for each wavelength separately. The relative PMT efficiencies $\epsilon_{j}$ are extracted separately in a final step from the ratio between the data and model occupancies, after all the other model parameters are characterized.

Systematic errors are introduced through uncertainties in the calibration variables, in particular those related to the laserball position, light distribution and wavelength. The optical fit is repeated with shifts applied to each calibration variable, and the output parameters are used to calculate the systematic change in the nominal fit results. The main systematic error comes from the laserball position uncertainties obtained by comparing the position provided by the manipulator hardware with a fitted position from the data. The main correction to the observed occupancy is the solid angle correction, which is inversely proportional to the square of the source-PMT distance (discussed in more detail in Section \ref{section:dataSeletion}). Consequently, even small deviations in the laserball position can create big variations in the corrected occupancy, affecting primarily the attenuation coefficients.

\subsection{Cross-check analysis of the media attenuations}
\label{subsec:crossCheck}

The attenuations of the inner detector medium, extracted from the main analysis fit, can be validated by a simplified and independent analysis of the laserball data. This independent analysis makes use of the calibration data with the laserball placed in different internal positions along a diagonal line passing through the center of the detector, and only considers the occupancies of two small groups of PMTs centered around the point where the diagonal line intersects the PMT support structure. Choosing the PMTs over a straight line ensures that the incidence angle and the angular distribution do not change, to first order, from PMT to PMT (the photons travel normal to the acrylic and the PMTs), leaving the attenuation as the main parameter in the optical model. 

The ratio of occupancies between two opposite PMTs (one in each side of the detector), in a run \emph{i}, can be modeled as:
\begin{equation}
\label{eq:ratio0}
\centering	
\frac{O_{i1}}{O_{i2}}=\frac{N_i\Omega_{i1}R_{i1}T_{i1}L_{i1}\epsilon_{1}e^{\left(-\sum_{k}d_{i1}^k\alpha_k\right)}}{N_i\Omega_{i2}R_{i2}T_{i2}L_{i2}\epsilon_{i2}e^{\left(-\sum_{k}d_{i2}^k\alpha_k\right)}}\:.
\end{equation}

Because the PMTs are aligned, one can assume that the distance traveled by light in the acrylic and in the external water is the same for each side ($d^a_{i1}=d^a_{i2}$ and $d^{w_{ext}}_{i1}=d^{w_{ext}}_{i2}$), yielding:
\begin{equation}
\label{eq:ratio1}
\centering	
\frac{O_{i1}}{O_{i2}}= \frac{\Omega_{i1}R_{i1}T_{i1}L_{i1}\epsilon_{1}}{\Omega_{i2}R_{i2}T_{i2}L_{i2}\epsilon_{2}}e^{-(d^{w_{\text{int}}}_{i1}-d^{w_{\text{int}}}_{i2})\alpha_{w_{\text{int}}}}\:.
\end{equation}

The ratio of the occupancies of the two opposite PMTs will, therefore, vary exponentially with the difference between the light paths inside the AV for each PMT, with a slope equal to the attenuation coefficient of the medium inside the acrylic vessel. Because the solid angle and the Fresnel transmission coefficients can be calculated numerically, they are fixed in this analysis. This leaves on the right side of equation \ref{eq:ratio1} a dependence on the distances $d^{w_{\text{int}}}_{i}$ as independent variables, and the internal water attenuation as the parameter to measure. The angular response and efficiency of the PMTs and the laserball light distribution, to first approximation, can be considered as constants.

\subsection{Method for measuring the group velocity of light in water}

The accurate knowledge of the group velocity of light is essential for the simulation, reconstruction and analysis of SNO+ data, since all conversions between photon travel times and travel distances rely on this parameter. The group velocity is given by the derivative of the angular frequency $\omega$ with respect to the wave number $k$, i.e. $v_g = \frac{d\omega}{dk}$. The group velocity of light with wavelength $\lambda$ in a medium with refractive index $n$ can be expressed as follows:
\begin{equation}
v_g = \frac{c}{n}\left(1+\frac{\lambda}{n}\frac{dn}{d\lambda}\right)\:.
\end{equation}

The group velocity of light in water was measured with laserball data in the water phase, making use of positions along the detector vertical axis. The measurements served as validation for the values used by the SNO+ simulation and reconstruction.

\begin{figure}[htbp]
	\centering
	\includegraphics[width=0.45\textwidth]{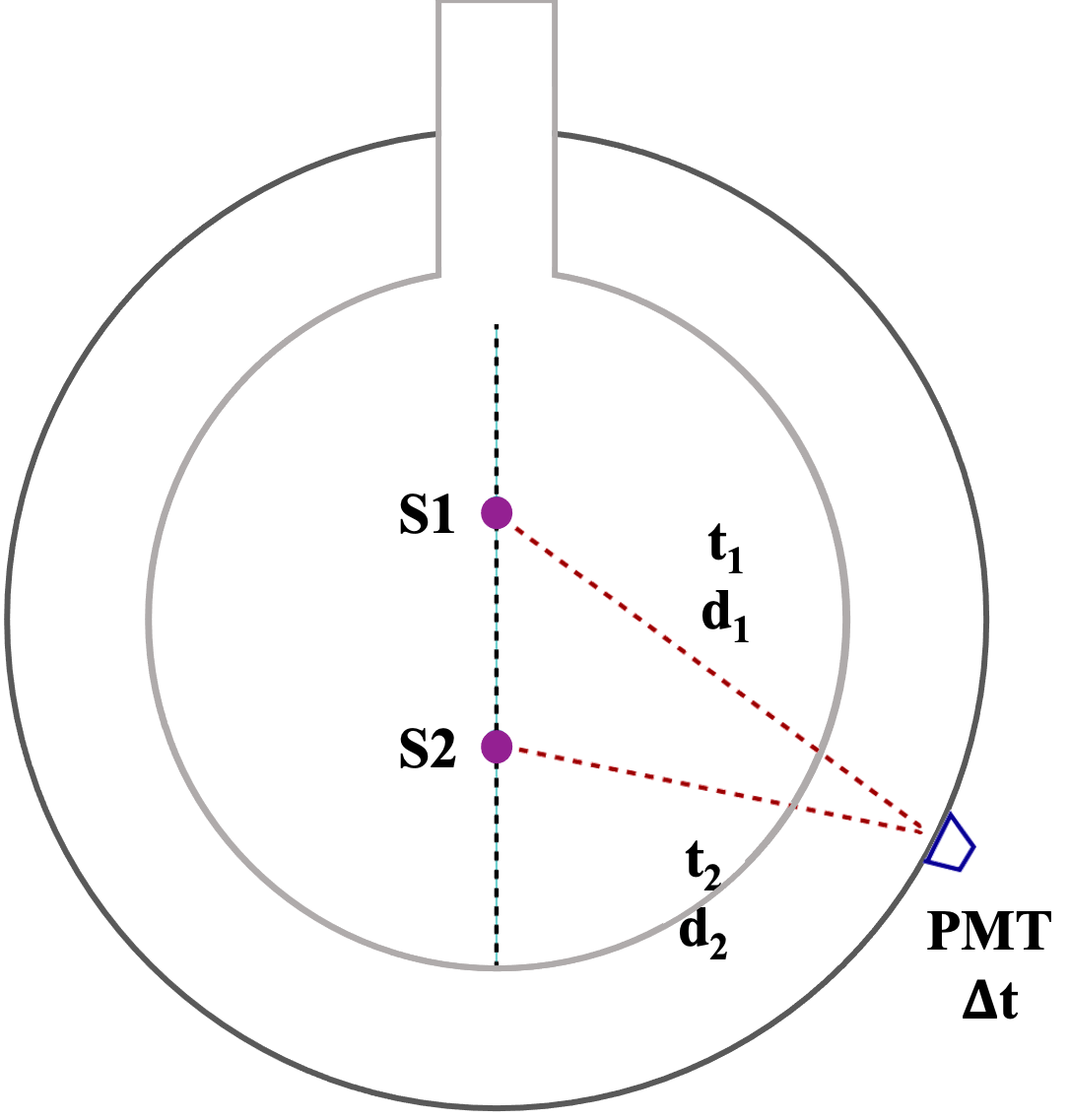}
	\caption{\label{fig:groupVel} Scheme of the group velocity measurement method.}
\end{figure}

The method relies on a PMT-by-PMT comparison of the prompt peak centroid from measured hit times between pairs of runs with the laserball in different positions. This comparison is done for a large number of PMTs, and employing several pairs of runs, at different vertical distances from each other. The group velocity is calculated from the differences between the times ($t_1$, $t_2$) and distances ($d_1$, $d_2$) for each PMT and run pair as:
\begin{equation}
v_g = \frac{d_1-d_2}{t_1-t_2}\:.
\end{equation}
The use of the same PMT from different runs makes this method independent of the PMT channel offset calibrations, that do use the knowledge of $v_g$, and is also much less sensitive to systematic uncertainties in the source position. The distances between the source position and the PMT, illustrated in Figure \ref{fig:groupVel}, are calculated assuming a straight line and by taking the source positions directly from the manipulator hardware. Grouping the PMT/run pairs according to the difference in distance between the source positions (S1 and S2) has shown that the group velocity was consistent across a wide range of distances.

\section{Water phase calibration data}
\label{section:data}

During the SNO+ water phase there were two main laserball data-taking campaigns: an internal laserball scan in December 2017, and internal and external scans in July 2018. During the December 2017 campaign, data were collected in a total of 31 internal positions (including four central positions with the laserball at different azimuthal orientations, to help understanding the anisotropies in its light output), for the six available wavelengths. This campaign had the main goal of commissioning the laser and laserball hardware, and the data were used to exercise the calibration data processing and analysis tools.

\begin{figure}[htbp]
	\centering
	\includegraphics[width=.8\textwidth]{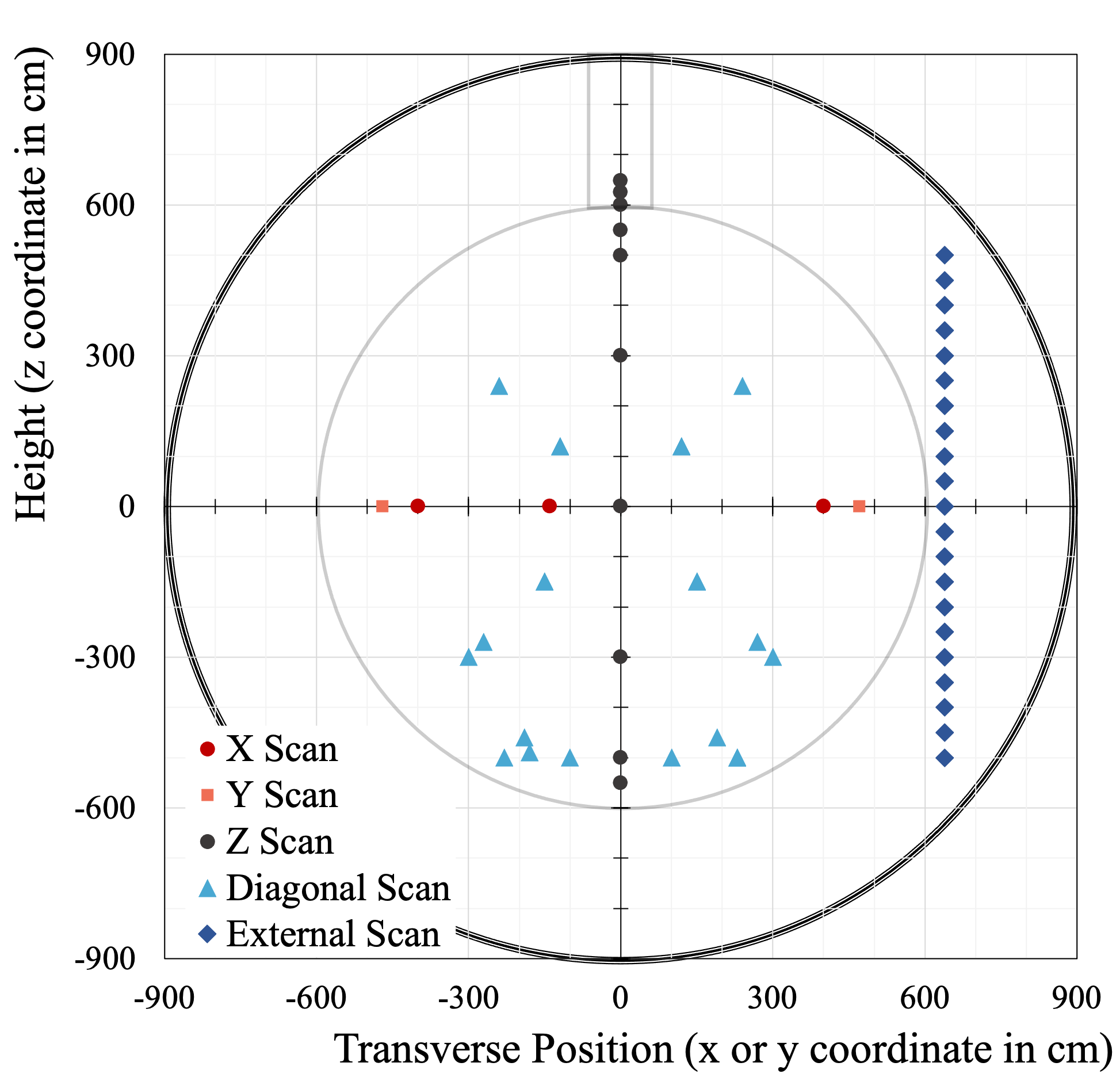}
	\caption{\label{fig:lbpositions} Laserball positions during the 2018 calibration campaign, projected in the transverse plane.}
\end{figure}

Similarly, during the July 2018 campaign data were collected at 42 internal positions, including the four central positions with different laserball orientations (Figure \ref{fig:lbpositions}). Additionally, data were collected at 19 positions along a vertical axis outside the AV. Each run (internal or external) had around 8700 online, inward facing PMTs. 
Since the data from the central positions are used as normalization in the analysis, these runs were typically one hour long for one of the laserball orientations, and 30 minutes for the other orientations. The run length of the off-center positions was 15 minutes. 

The laser emitted 40 light pulses per second, and the intensity was kept low using neutral density filters so that only about 5\% of the PMTs register hits for each laser pulse. This way, we ensured that the corrections applied to account for multiple photons hitting a single tube were small. The stability of the laser emission was monitored live during the calibration campaigns by looking at the integrated number of hit PMTs per second. Afterwards, the stability of the data was evaluated offline during the data quality stage of the analysis. The indicators used were the number of hit PMTs per laser pulse, as well as the width of this distribution over the course of a laserball run. Fluctuations in the number of PMTs registering a hit per laserball pulse were smaller than 2\%.
 
 In November 2017 the $^{16}$N source was deployed inside the AV, and data were collected in 80 different positions, along the detector horizontal, and vertical axis. The runs were between 20 and 30 minutes long, with an average rate of 30 to 60 tagged events per second, depending on the settings of the DT generator supply, like the CO$_2$ gas flow rate.

 \subsection{Laserball data selection and cuts}
 \label{section:dataSeletion}

\begin{figure}[htbp]
	\centering
	\includegraphics[width=1\textwidth]{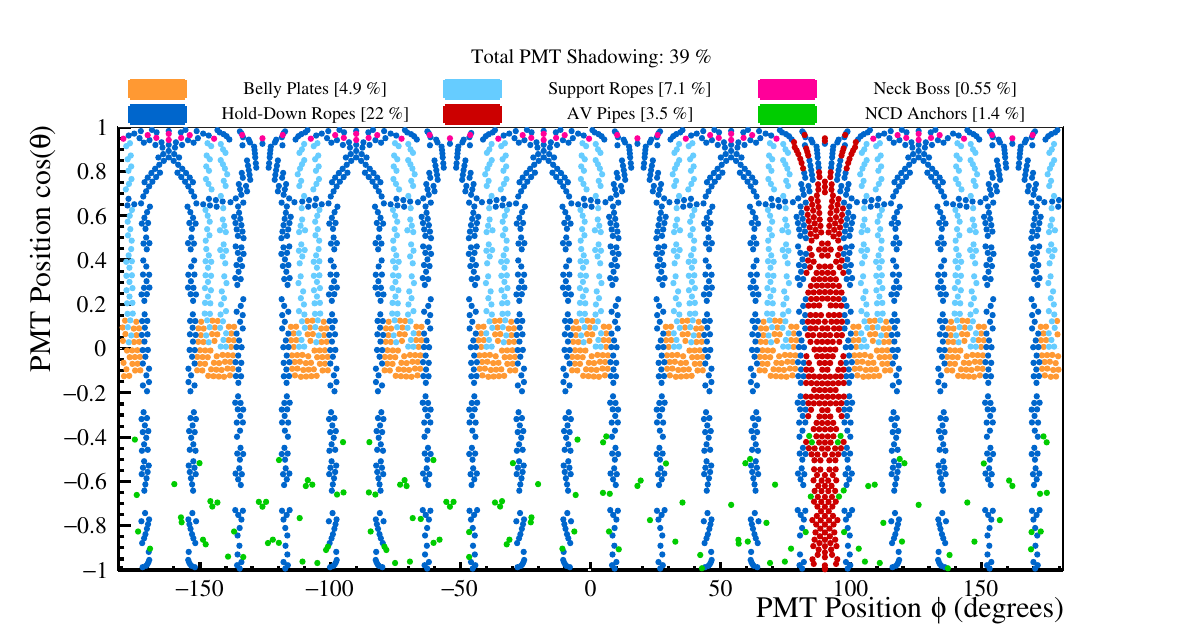}
	\caption{\label{fig:centralShadowing} Map of PMTs shadowed by detector components for a central laserball position. The PMTs are considered shadowed if their light path, starting at the source position, comes within a tolerance distance or intersects the detector components not included in the optical model. The tolerance distance is 30 cm to the AV belly plates, and 15 cm to all the other components.}
\end{figure}

The occupancy of each online, inward facing PMT in each run is a candidate data point for the optical fit, giving approximately 4$\times 10^{5}$ data points that enter the fit for each wavelength. The analysis cut that results in the biggest fraction of PMTs excluded from the data set is the "shadowing" cut. It removes PMTs whose light paths are within a tolerance distance or intersect detector components not included in the optical model, such as the AV support ropes and AV pipes. Figure \ref{fig:centralShadowing} shows a map of the PMTs shadowed by detector components for a central laserball position. This shadowing cut results in 39\% of the data points being removed in the normalization run. Of the remainder, up to 28\% of the data points were excluded from each internal off-axis run, depending on the position, by applying the same cut. 

\begin{figure}[htbp]
	\centering
	\includegraphics[width=.95\textwidth]{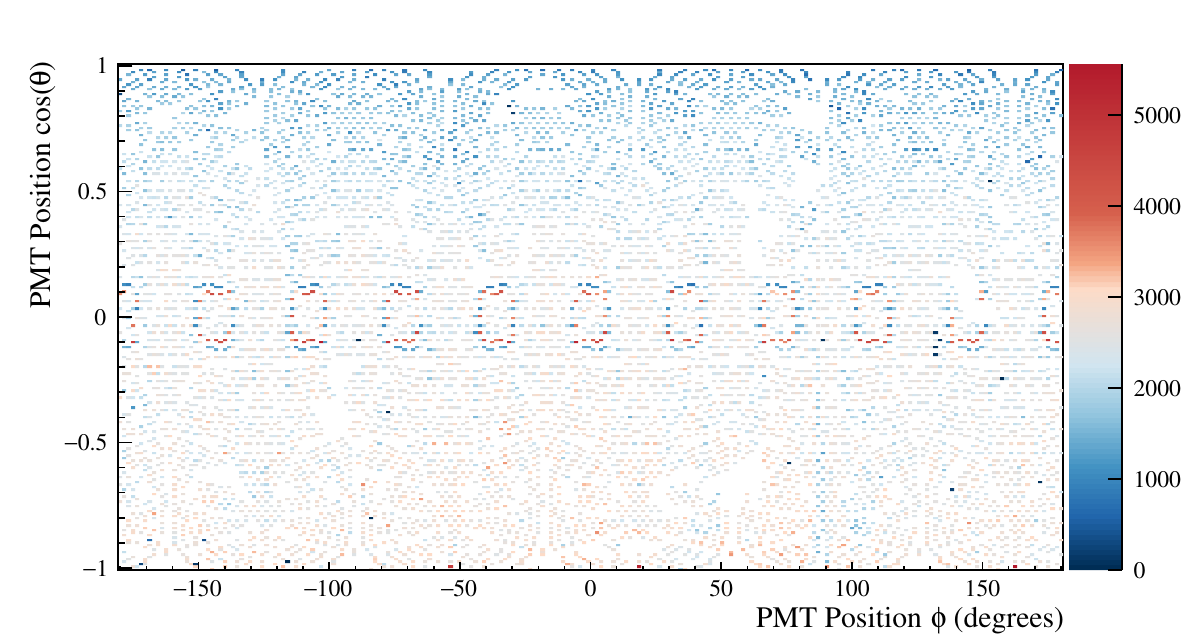}
	\includegraphics[width=.95\textwidth]{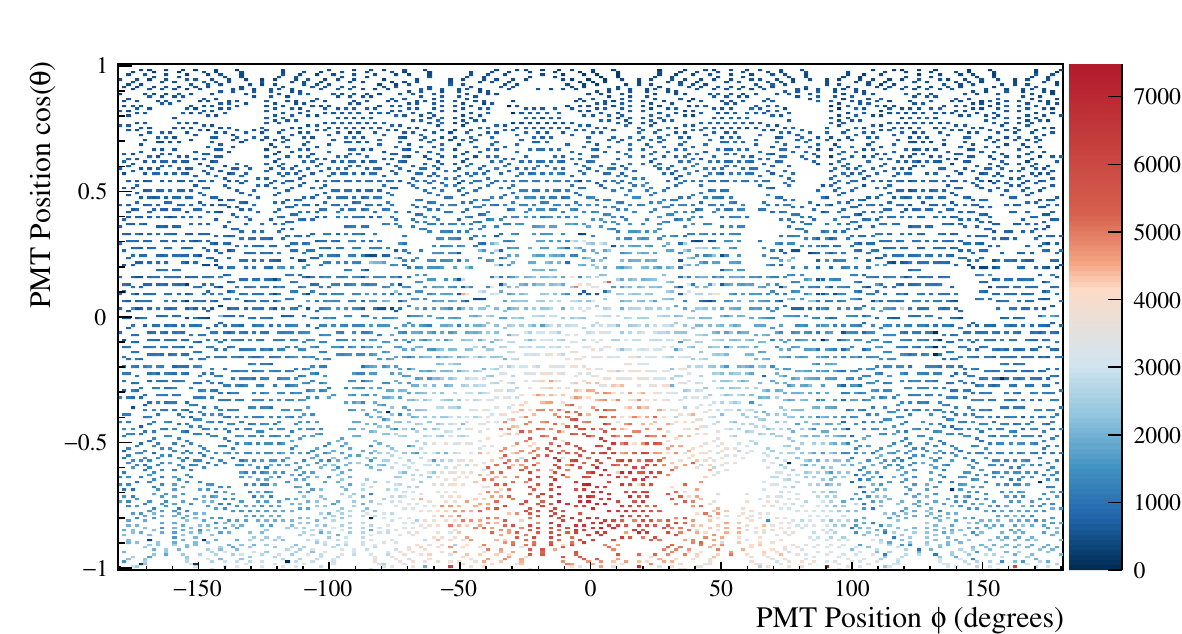}
	\caption{\label{fig:dataOccMaps} Map of PMT occupancies for central (top) and off-center (bottom) laserball runs from the 2018 calibration campaign.}
\end{figure}

Figure \ref{fig:dataOccMaps} shows maps of the PMT occupancies for a central and an off-center laserball run, prior to any analysis cuts. For the central run, it is possible to observe directly in the data the shadowing caused by the detector components: circles of lower occupancy PMTs around the detector equator, shadowed by the rope loops inside acrylic panels mounted on the outside of the AV ("belly plates"). The shadowing effects by the hold-down rope net on the top of the AV are also clearly visible, as is the laserball hardware shadowing, resulting in lower occupancy PMTs in the top part of the detector. Such effects are harder to observe in the raw data of the off-center run. In the latter case, the PMTs closer to the laserball will have an occupancy about 15--16\% larger than the ones in the opposite side of the detector, mostly due to the solid angle effect.

The variable that affects the solid angle calculation the most is the source position. The solid angle is proportional to $1/R^2$, where $R$ is the radial position of the laserball. For example, a radial position scale factor of 1.01 changes the solid angle correction for a given PMT by about 2\%. Since the variations of the occupancy due to the solid angle are larger than those due to the optical parameters, the laserball position needs to be determined with a high level of accuracy. Although the manipulator system provides an estimate of the laserball position, its positioning algorithm is based on the length of the ropes that move the calibration source, which depends on the rope tension and is therefore not precise enough when moving it to positions outside the vertical axis.

Comparing the position fitted from the laserball data to the manipulator position provides the systematic variation to be considered in the main analysis fit. The agreement between the fitted position and the manipulator was better than 2 cm for central positions, and ~4 cm for high radius positions. A laserball position uncertainty of 4 cm was used when calculating the systematic uncertainties of the optical model parameters.

The introduction of the external laserball data in the optical calibration fit was a new feature and improvement of the analysis, relative to SNO. Nevertheless, the only external data points considered for the analysis came from PMTs whose light paths from the source were fully contained in the external water region. This selection was made to avoid uncertainties in the solid angle calculation for PMTs that would see light crossing the full AV (and intersecting the acrylic boundaries twice).

\begin{figure}[htbp]
	\centering
	\includegraphics[width=.4\textwidth]{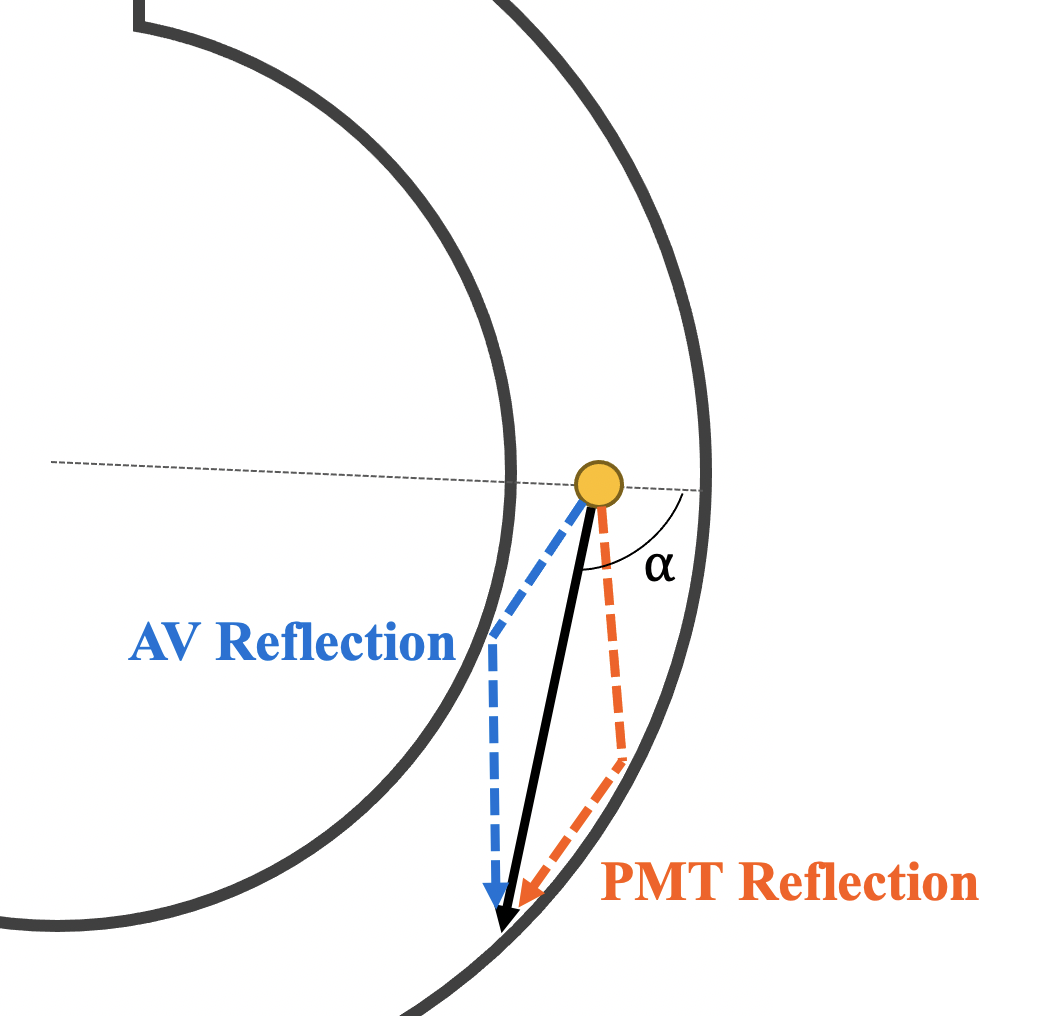}
	\includegraphics[width=.59\textwidth]{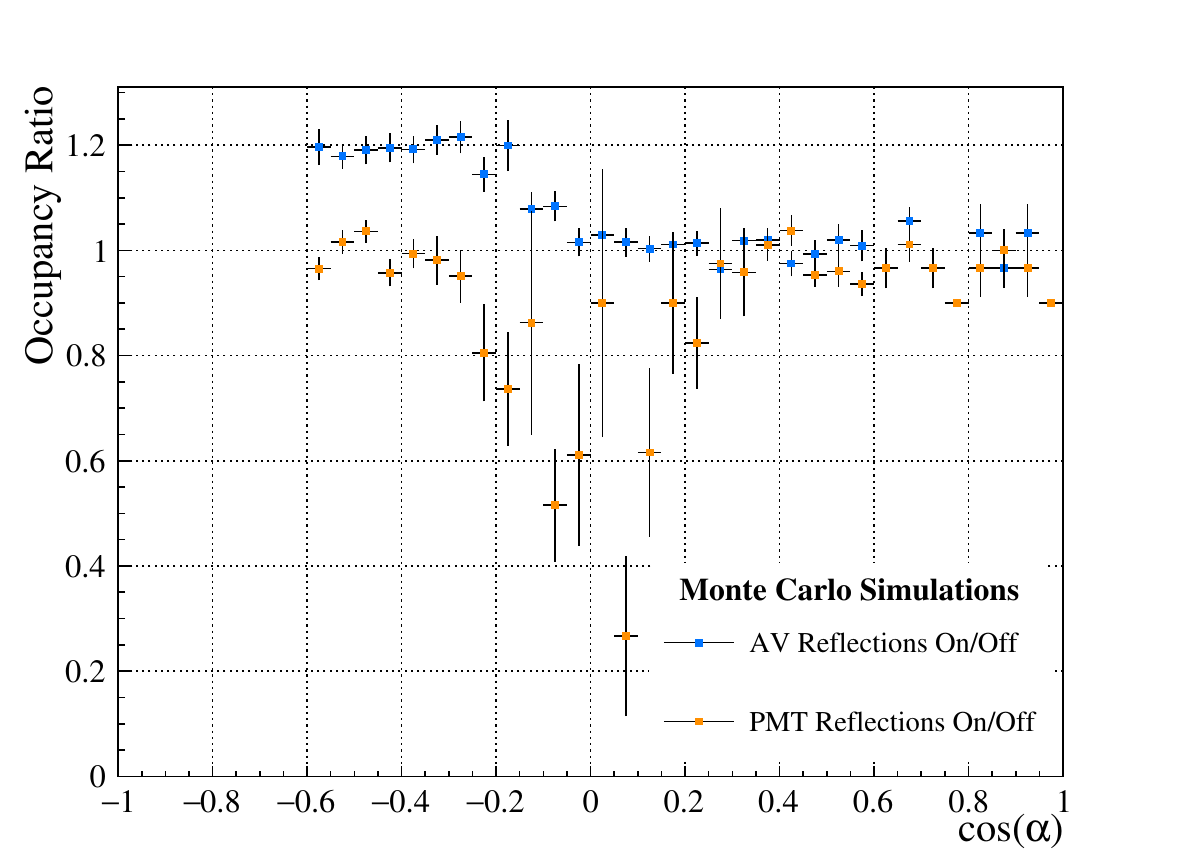}
	\caption{\label{fig:externalReflections} Ratio of PMT occupancies in MC simulations with AV reflections on and off (blue) and PMT reflections on and off (orange), as a function of cos($\alpha$), where $\alpha$ is the angle between the laserball position vector and the vector pointing from the center of the laserball to the PMT.}
\end{figure}

In addition, several cuts had to be implemented in order to deal with PMTs whose measured occupancy was affected by light reflected from the AV outer surface or other PMTs. These cuts were determined by comparing laserball simulations with and without reflections from the AV and from the PMTs. Figure \ref{fig:externalReflections} shows the ratio of occupancies for each PMT between the simulations with reflections on and with reflections off, as a function of cos($\alpha$), where $\alpha$ is the angle between the vector pointing from the detector center to the laserball position, and the vector pointing from the center of the laserball to the PMT. 

PMTs further away from the laserball will have a 20\% overestimated occupancy due to light reflected from the AV surface that reaches the PMT in the 8 ns prompt time window. Furthermore, light entering the PMT reflector assembly at a given angle can be reflected and not detected by the PMT. Both these effects are not accounted for in the optical model (Equation \ref{eq:optmodel}), and the comparisons of simulation with data for each external positions were used to determine cos($\alpha$) cuts to exclude the affected PMTs. Due to the strict light path type selection and PMT cuts for the external runs, between 94\% and 97\% of the number of data points from each external position was excluded from the fit. Even though this results in a small sample of about 435 data points for each external position, the fit is performed using all the data available from the 61 internal and external positions, which provides enough statistics to estimate the optical parameters and mitigates any biases due to PMT sampling.


\section{Results of the optical calibration analysis}
\label{section:results}


The parameters of the optical model presented in Section \ref{section:method} were extracted from the $\chi^2$ minimization using the internal and external laserball data. The minimization assumed the same water attenuation coefficients for the internal and external water. This decision was made after performing the fit with the attenuations separated, and verifying that the measured external water attenuation coefficients were compatible with the ones for the internal water, but with much larger uncertainties. The combined internal and external water attenuation coefficients measured in this analysis are presented on the left side of Table \ref{tab:AttValues}. Adding the external laserball data to the analysis allowed to perform the first \textit{in situ} measurement of the effective acrylic vessel attenuation coefficients, shown on the right side of Table \ref{tab:AttValues}.

\begin{table}[htbp]
	\caption{Fitted water attenuation coefficients, $\alpha_w$, and effective acrylic attenuation coefficients, $\alpha_a$, and their corresponding statistical and systematic uncertainties.}
	\label{tab:AttValues}
	\centering
	\resizebox{\textwidth}{!}{\begin{tabular}{ | c || c | c | c || c | c | c | }
			\hline
			$\lambda$ &  $\alpha_{\textrm{w}} $ & $\sigma_{\textrm{stat}}$ & $\sigma_{\textrm{syst}}$ & $\alpha_{\textrm{a}} $ & $\sigma_{\textrm{stat}}$ & $\sigma_{\textrm{syst}}$ \\ 
			{(nm)} & ($\times 10^{-5}$ mm$^{-1}$) & ($\times 10^{-5}$ mm$^{-1}$) & ($\times 10^{-5}$ mm$^{-1}$) & ($\times 10^{-3}$ mm$^{-1}$) & ($\times 10^{-3}$ mm$^{-1}$) & ($\times 10^{-3}$ mm$^{-1}$) \\ \hline
			337 & 1.331 & 0.006 & 0.489 & 9.19 & 0.05 & 1.12 \\
			365 & 1.013 & 0.005 & 0.421 & 4.31 & 0.04 & 0.86 \\
			385 & 0.859 & 0.005 & 0.431& 3.15 & 0.04 & 0.84 \\
			420 & 0.819 & 0.005 & 0.423 & 2.61 & 0.04 & 0.81 \\
			450 & 0.943 & 0.005 & 0.419 & 2.75 & 0.04 & 0.81 \\
			500 & 2.615 & 0.005 & 0.443 & 2.43 & 0.04 & 0.83 \\
			\hline
	\end{tabular}}
\end{table}

\begin{figure}[htbp]
	\centering
	\includegraphics[width=.9\textwidth]{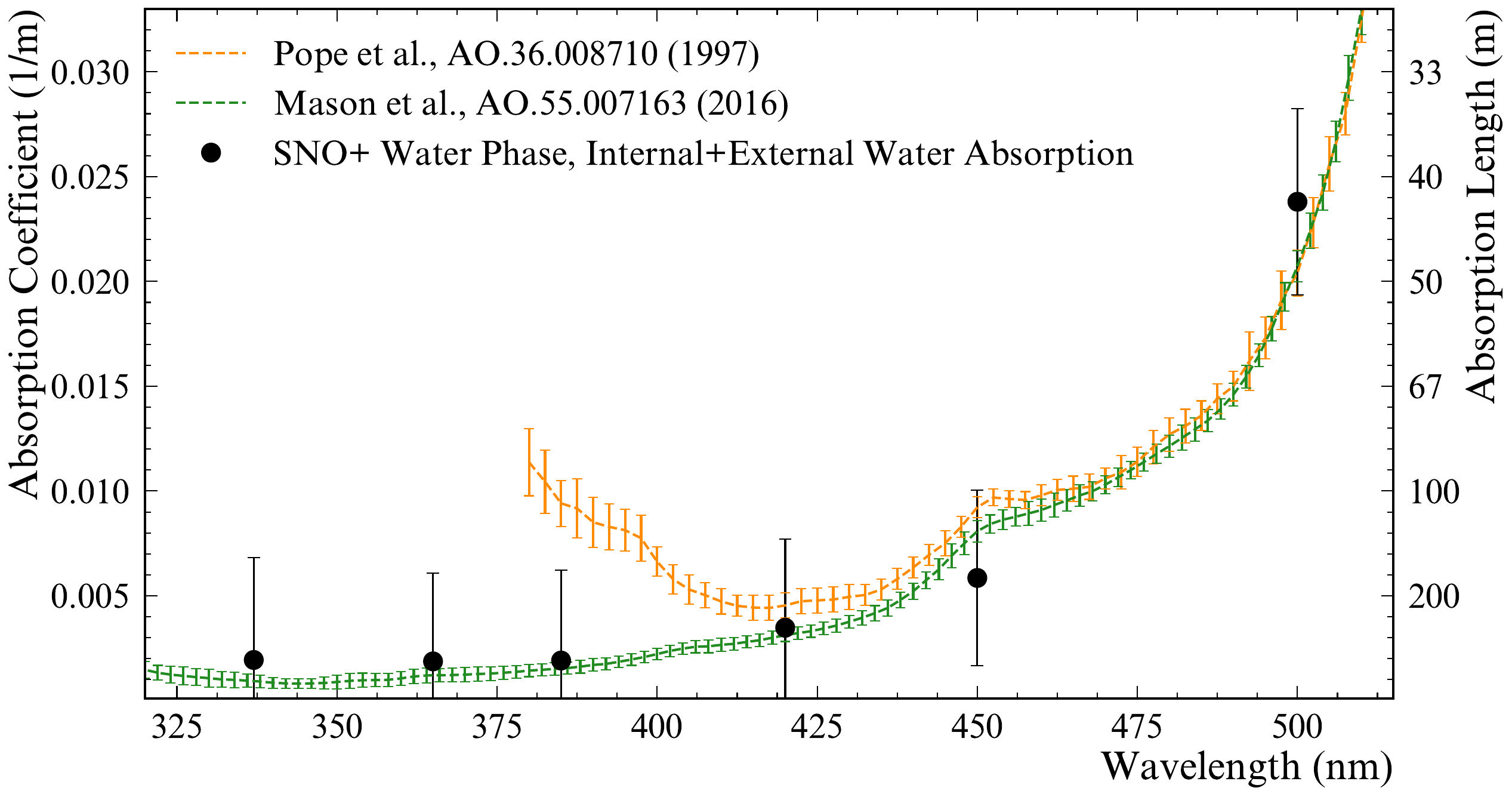}
	\caption{\label{fig:waterAbs} Internal and external water absorption coefficients (left axis) and lengths (right axis) as a function of wavelength. Shown are the results from the Optical Calibration Analysis for the data of the July 2018 laserball internal and external scans (black), after correcting the measured prompt attenuation for the effects of the Rayleigh scattering. The orange and green lines are water absorption values from \cite{pope} and \cite{mason}, respectively.}
\end{figure}

The water attenuation coefficients, $\alpha_{\textrm{w}}$, include the effects of light absorption, $\alpha_{\textrm{abs}}$, and of Rayleigh scattering, $\alpha_{\textrm{RS}}$. The Rayleigh scattering is responsible from removing light from the prompt peak, i.e. typically scattered light will arrive at the PMTs at later times. However, there is a probability that light will be scattered forward and be detected as prompt light. The forward scattered light will increase the prompt attenuation length $\lambda_{\textrm{w}}$, thus decreasing the attenuation coefficient $\alpha_{\textrm{w}}=1/\lambda_{\textrm{w}}$. Therefore, the forward-scattered light reduces the contribution of the Rayleigh scattering coefficient to the attenuation coefficient by a factor $k$. The relation between the three coefficients is then:
\begin{equation}
\alpha_{\textrm{w}} = \alpha_{\textrm{abs}} + k\:\alpha_{\textrm{RS}}\:.
\end{equation}
If 100\% of the scattered light is removed from the prompt peak, $k$ should be 1. If 100\% of the scattered light is forward scattered and is included in the prompt peak, then $k = 0$ and the attenuation only includes the absorption effect. The value of $k$ considered in this analysis was 0.82, obtained from studies of the fraction of light removed from the prompt peak due to scattering conducted in SNO \cite{moffat}. Because the Monte Carlo simulation must model both absorption and scattering, the scattering contribution is subtracted from the measured coefficient, and the resulting absorption coefficient is used as an input to the Monte Carlo.\footnote{The Rayleigh scattering coefficients used were determined by the fixed calibration system of optical fibers.} Figure \ref{fig:waterAbs} shows the water absorption coefficients, which are in good agreement with literature values from \cite{pope,mason}. 

\begin{figure}[htbp]
	\centering
	\includegraphics[width=.87\textwidth]{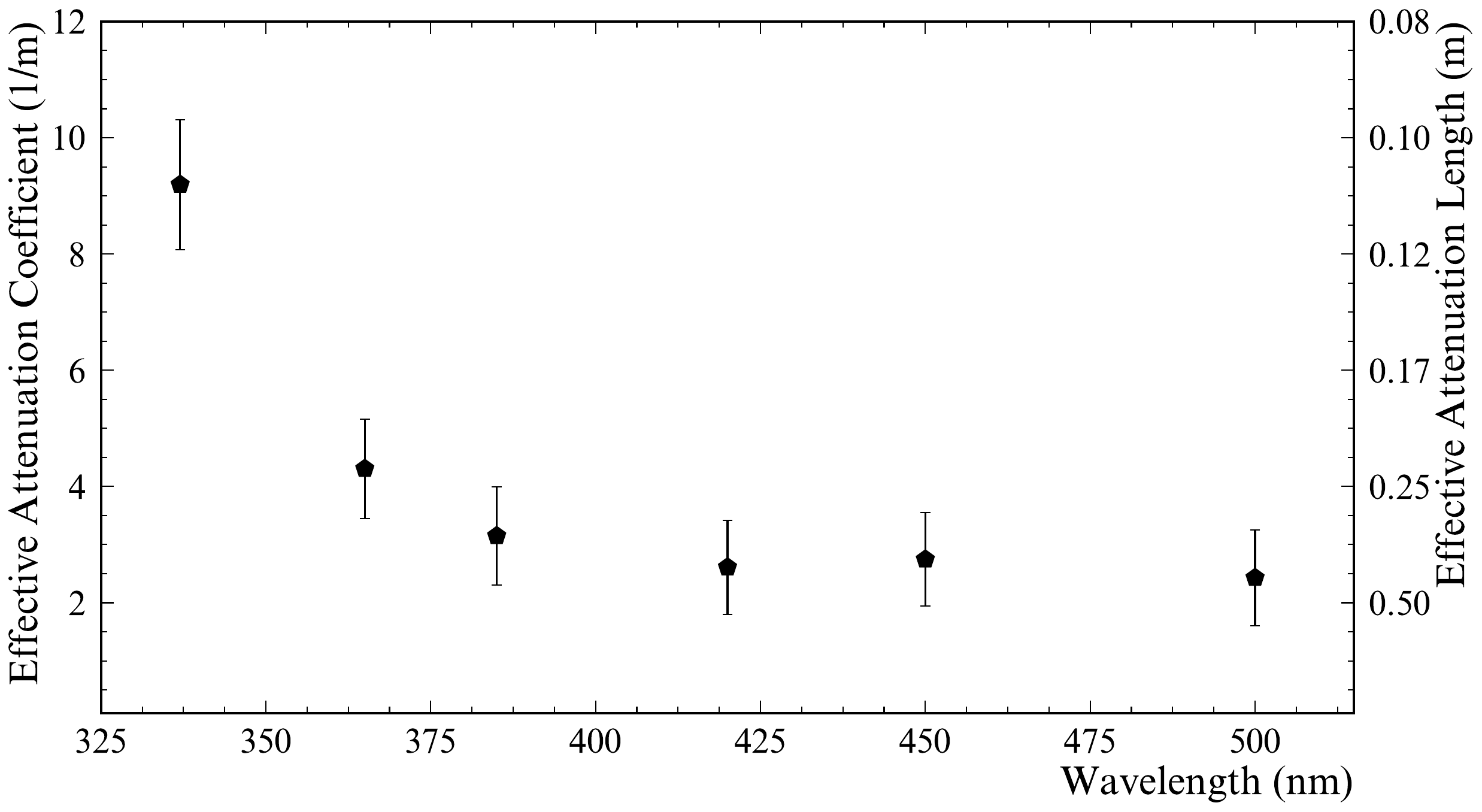}
	\caption{\label{fig:acrylicAtt} Effective acrylic vessel attenuation coefficients (left axis) and lengths (right axis) as a function of wavelength. The results come from the Optical Calibration Analysis of the data of the 2018 laserball internal and external scans. These are the first \textit{in situ} measurements of the effective acrylic vessel attenuation.}
\end{figure}

In the case of the acrylic vessel, the fitted effective attenuation coefficients, shown in Figure \ref{fig:acrylicAtt}, were directly propagated to the SNO+ Monte Carlo as absorption lengths. We model the acrylic attenuation measurements as effective bulk transmission, since the Monte Carlo assumes that the acrylic is uniform. However, other bond or surface-related effects cannot be excluded when interpreting these results. 

\begin{figure}[htbp]
	\centering
	\includegraphics[width=.85\textwidth]{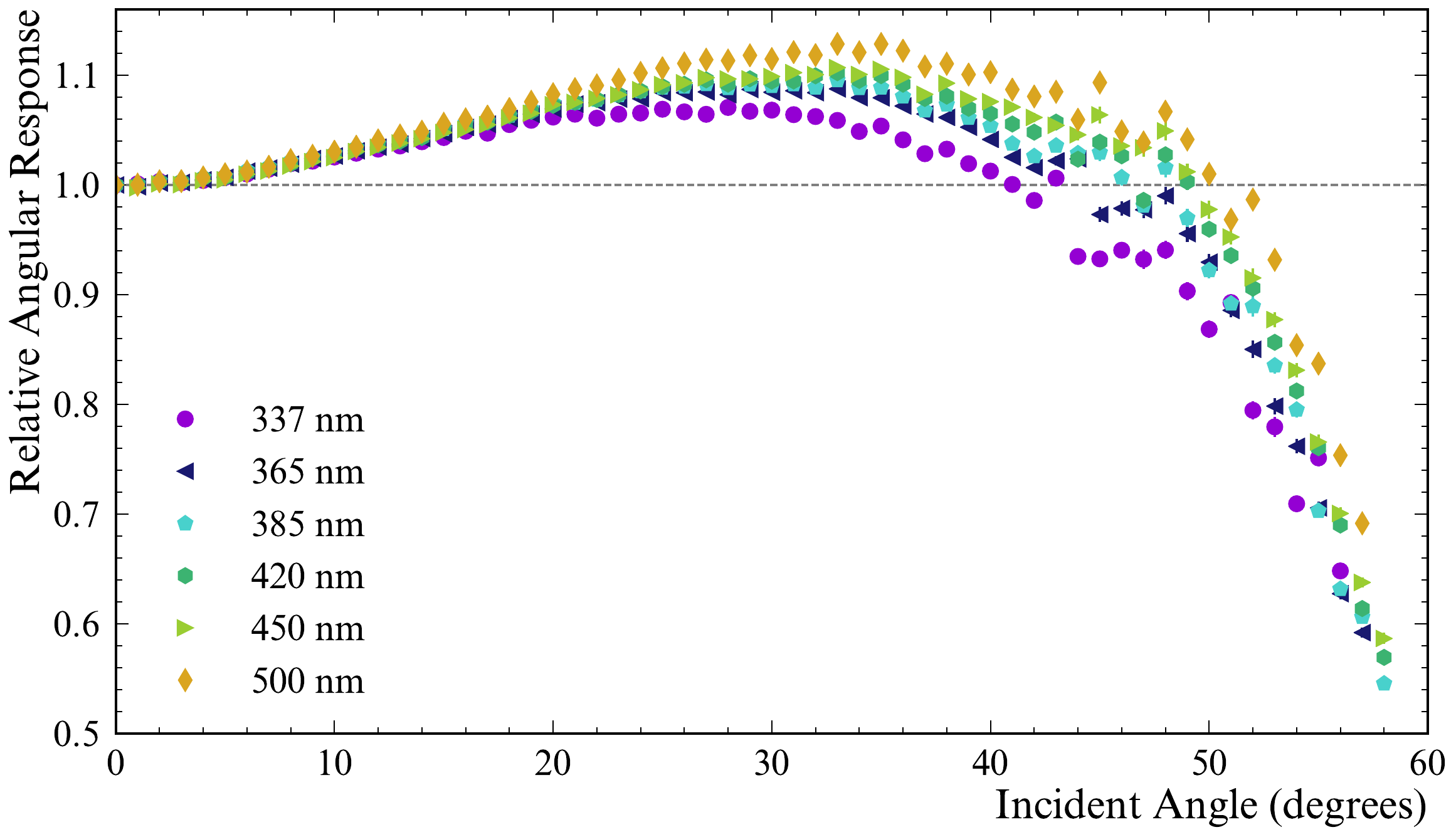}
	\caption{\label{fig:pmtResponse} Relative PMT-concentrator assembly angular response as a function of the incidence angle, for the six laserball wavelengths used during the 2018 internal and external scans. The angular response values are normalized to the one at a normal incidence (0 degrees). The inclusion of the external scan data allowed for values above 45 degrees to be measured \textit{in situ} for the first time. Only the statistical uncertainties are displayed.}
\end{figure}

In addition to the attenuation lengths, the response of the PMTs and concentrators as a function of incidence angle was also measured. The angular dependence is parameterized as a simple binned response function, with bins in steps of 1 degree ranging from normal incidence (0 degrees) to the highest angle possible, where normal incidence is defined as normal to the front plane of the PMT and concentrator assembly. The internal scan positions are only able to cover an incident angle up to 45 degrees. The addition of the external laserball data allowed to measure the response at higher angles \textit{in situ} for the first time. Figure \ref{fig:pmtResponse} shows the PMT and concentrator assembly angular response for the six laserball wavelengths, normalized by the response to light at normal incidence. The concentrators are responsible for increasing the angular response with incidence angle up to a peak at 30 -- 35 degrees. However, beyond 45 degrees, light entering the PMT and concentrator assembly will be mostly reflected back out due to the design of the concentrators' shape. It is important to note that there is a strong correlation between the effective acrylic attenuation coefficients and the PMT angular response parameters at high angles, between 40 and 50 degrees. The measured angular responses are directly introduced in the SNO+ Monte Carlo as the grey disc PMT model absorption probabilities. 

\begin{figure}[htbp]
	\centering
	\includegraphics[width=.85\textwidth]{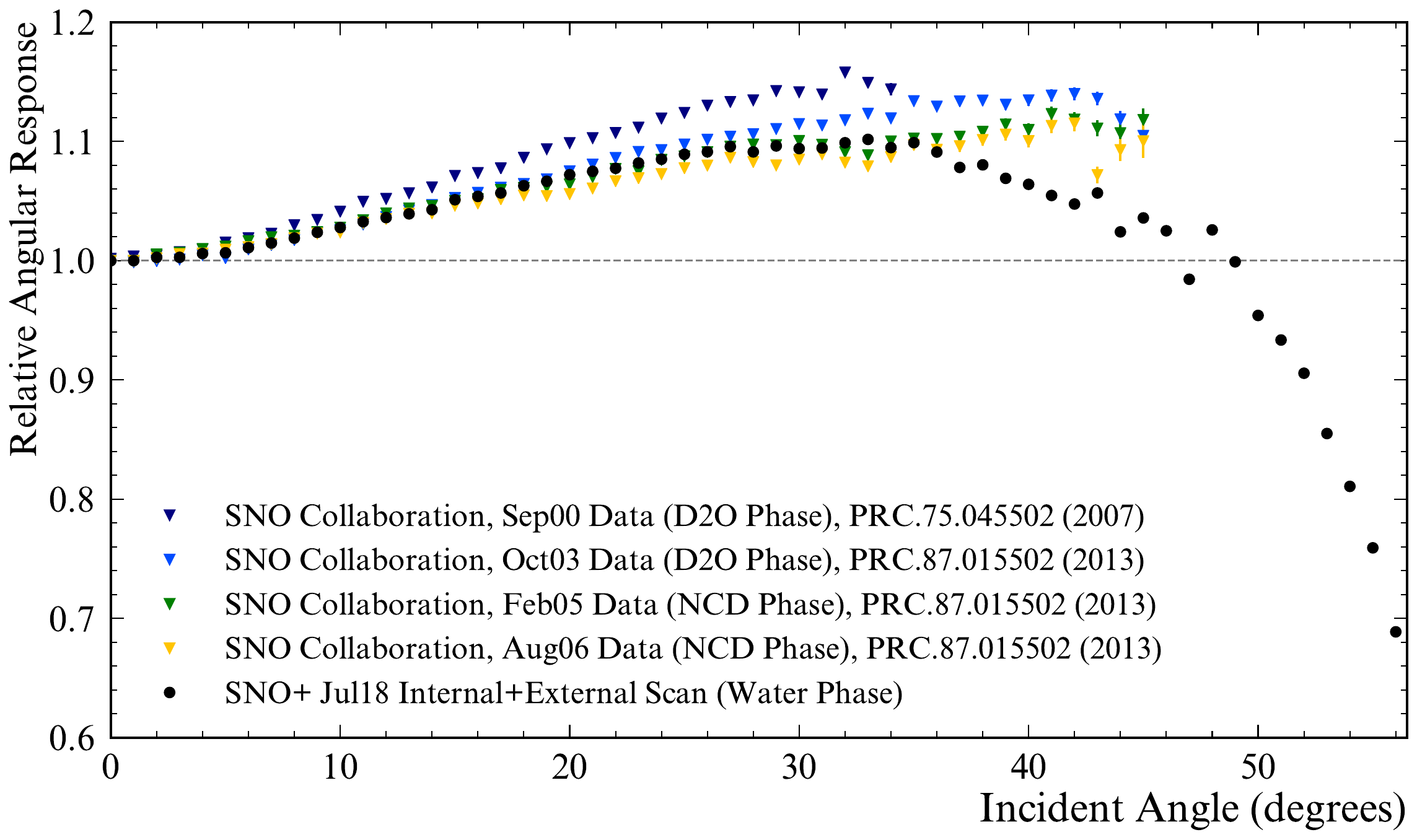}
	\caption{\label{fig:pmtResponseSNO} Relative PMT-concentrator angular response at 420 nm as a function of the incidence angle. In black are shown the measurements from the SNO+ water phase optical calibration analysis, compared with previous measurements from the SNO experiment. Only the statistical uncertainties are displayed.}
\end{figure}

Figure \ref{fig:pmtResponseSNO} compares the measured angular response at 420 nm with previous ex-situ measurements from SNO. It is important to notice that the angular response has been decreasing over time since the beginning of SNO, due to the degradation of the concentrator’s optical surface. This degradation, which made areas of the concentrators reflect more diffusely, has been directly observed in old concentrators that were removed and replaced with new ones during the SNO to SNO+ transition phase. The observed degradation does not seem to follow a pattern between PMTs, making it very difficult to create a model that would characterize its evolution with time.

\subsection{Measurement of the group velocity of light in water}

\begin{figure}[htbp]
	\centering
	\includegraphics[width=1\textwidth]{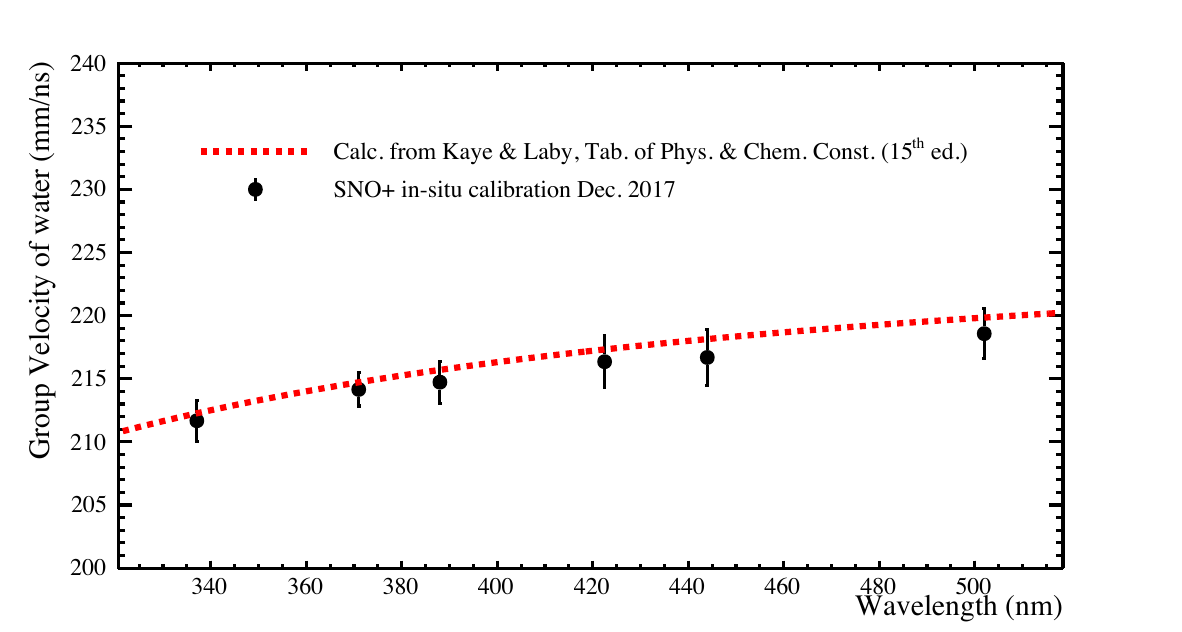}
	\caption{\label{fig:groupVelmeasurements} Group velocity of water as a function of wavelength, as measured \textit{in situ} in SNO+ with the December 2017 laserball data.	For comparison,	the	parameterization used in the SNO+ Monte Carlo and reconstruction, from \cite{groupV}.}
\end{figure}

The group velocity was measured using the data at each laserball wavelength. For this measurement, we used several runs taken along the vertical axis of the detector in December 2017, and selected only the data points for which the path difference between the source and each PMT, between the two runs, was the largest, to minimize the relative effect of position uncertainty. The results are shown in Figure \ref{fig:groupVelmeasurements}, and are consistent with the values used by the SNO+ Monte Carlo and reconstruction.

\subsection{Additional tuning of the detector model}

After propagating the optics measurements to the SNO+ Monte Carlo, there are two further aspects that need to be tuned: the collection efficiency scale factor, and the reflections of the PMT grey disc model that impact the late-light distributions. The collection efficiency scale factor was extracted by comparing the prompt light of data from the $^{16}$N in the center of the detector, with Monte Carlo simulations tuned with the optical analysis measurements. 

Tuning the grey disc model reflections included developing a parameterization for the reflections from the PMT-concentrator assembly. As shown in Figure \ref{fig:lightPaths_timeResid}, the time residual distribution for a central laserball run shows two prominent features produced by the PMTs: a specular reflection peak and an earlier peak coming from a preferred reflection mode named "35 degree PMT reflections", referring to the typical outgoing angle of photons with normal incidence. 

\begin{figure}[htbp]
	\centering
	\includegraphics[width=0.75\textwidth]{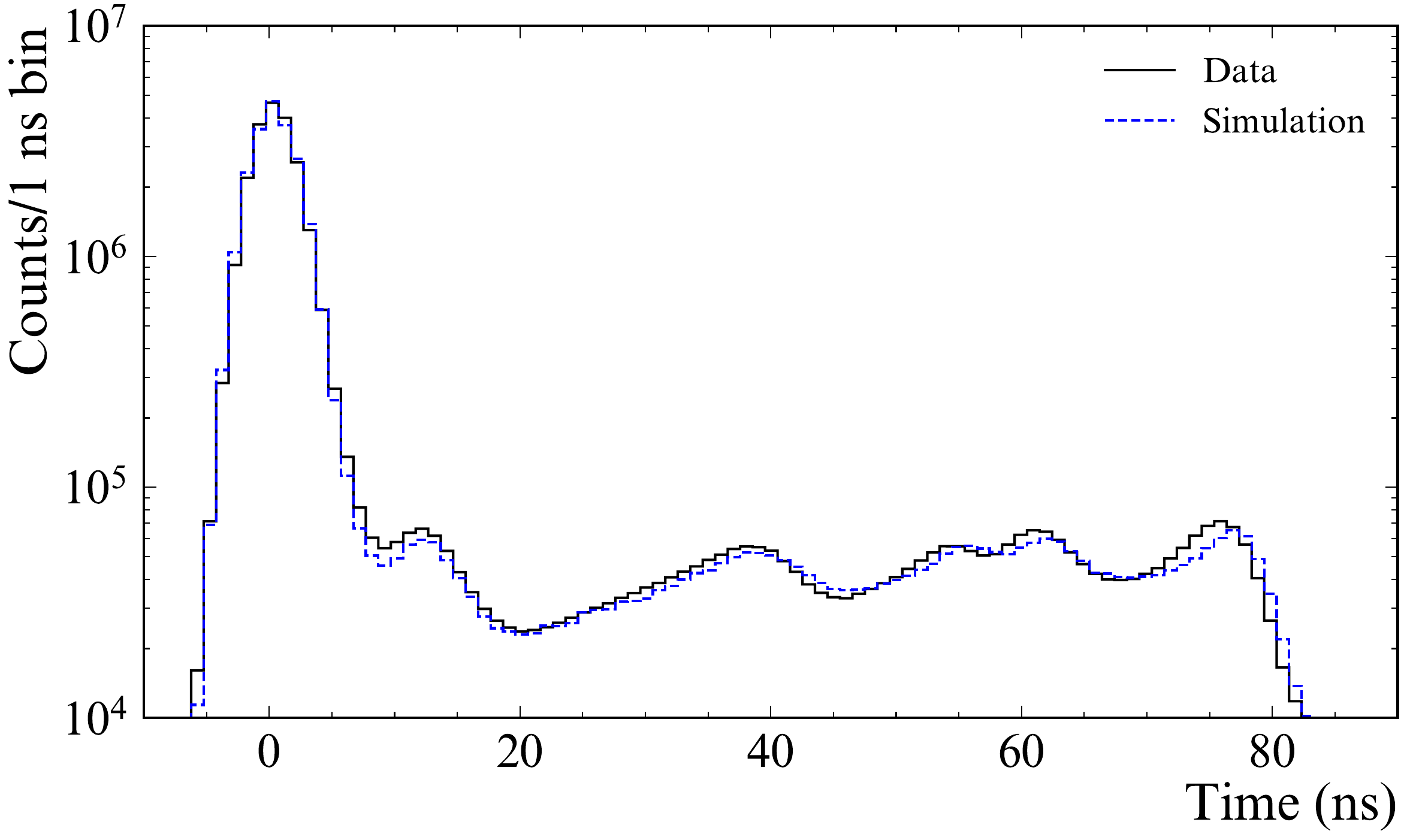}
	\caption{\label{fig:latelight} Time residual distribution for a central laserball run at 420 nm from data (black) and a simulation (blue) after tuning the SNO+ Monte Carlo with the measured optical parameters and adding the parameterization for the PMT reflection model.}
\end{figure}

The parameterization was done by, first, studying the outgoing angles of photons impacting the full 3D PMT model using simulations. The smear around the two reflection modes, as well as their evolution as a function of incident angle were encoded as free parameters in the parameterization model. The parameters were tuned to the data time residual distributions from 420 nm laserball runs at four radial positions from the center to the edge of the AV. This study also allowed to obtain reflection probabilities as function of incident angle, which are an input to the grey disc PMT model. Despite the simplicity of the reflection parameterization for the PMT-concentrator assembly, the late light distribution in the Monte Carlo shows a good agreement with the data, as can be seen in Figure \ref{fig:latelight}. 

\section{Validating the detector response model with the $^{16}$N source}
\label{section:n16validation}

\begin{figure}[htbp]
	\centering
	\includegraphics[width=.99\textwidth]{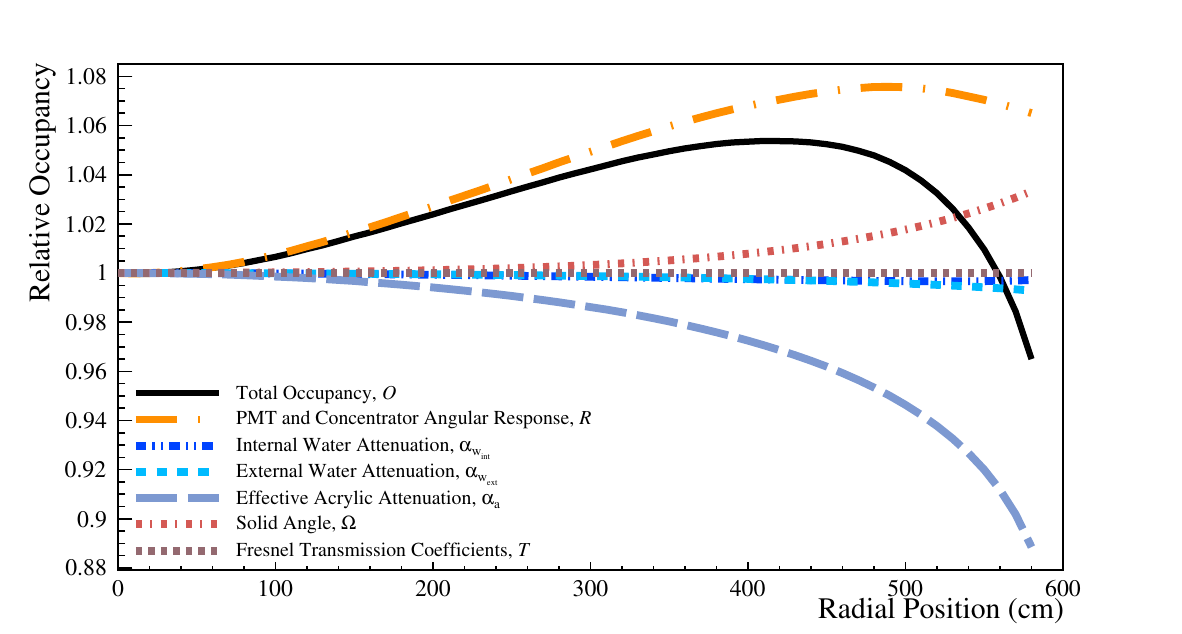}
	\caption{\label{fig:model} Contribution of the different optical model parameters to the integrated occupancy of all PMTs, as a function of radial position. The occupancy at each position is normalized by the occupancy at the center of the detector.}
\end{figure}

As discussed in Section \ref{section:motivation}, the optical properties of the SNO+ detector are responsible for variations of the energy response with radial position. This is illustrated in a simplified way by Figure \ref{fig:model}, which shows how each parameter of Equation \ref{eq:optmodel} independently affects the occupancy as a function of radial position. The curves are the sum of the calculated occupancy for all PMTs as a function of event radial position, divided by the summed occupancy for an event at the center. Comparing the total model occupancy (Equation \ref{eq:optmodel}), calculated using the optical calibration measurements at a single wavelength, with the occupancy curves calculated for each parameter gives an insight into which optical properties contribute the most for the overall detector response variations with position. Figure \ref{fig:model} shows that these are the PMT-concentrator assembly angular response and the effective acrylic attenuation. The variations of the detector response with position are one of the main contributors to the energy scale systematic uncertainty. 

After tuning the SNO+ Monte Carlo with the measured optical parameters, the detector model was validated by comparing the total number of hits ($N_\text{hit}$) created by $^{16}$N source events in data with simulations, at different positions inside the AV. The comparison using $N_\text{hit}$, instead of energy, avoids effects inherent to the event reconstruction. The $^{16}$N data selection criteria focused on prompt PMTs with time residual between -10 and 8 ns, which is the prompt time window used for energy reconstruction in SNO+ (prompt $N_\text{hit}$). The detector's state at the time of the $^{16}$N runs was also accounted for in this validation, by using only online channels, and with valid time and charge calibrations.

\begin{figure}[htbp]
	\centering
	\includegraphics[width=.8\textwidth]{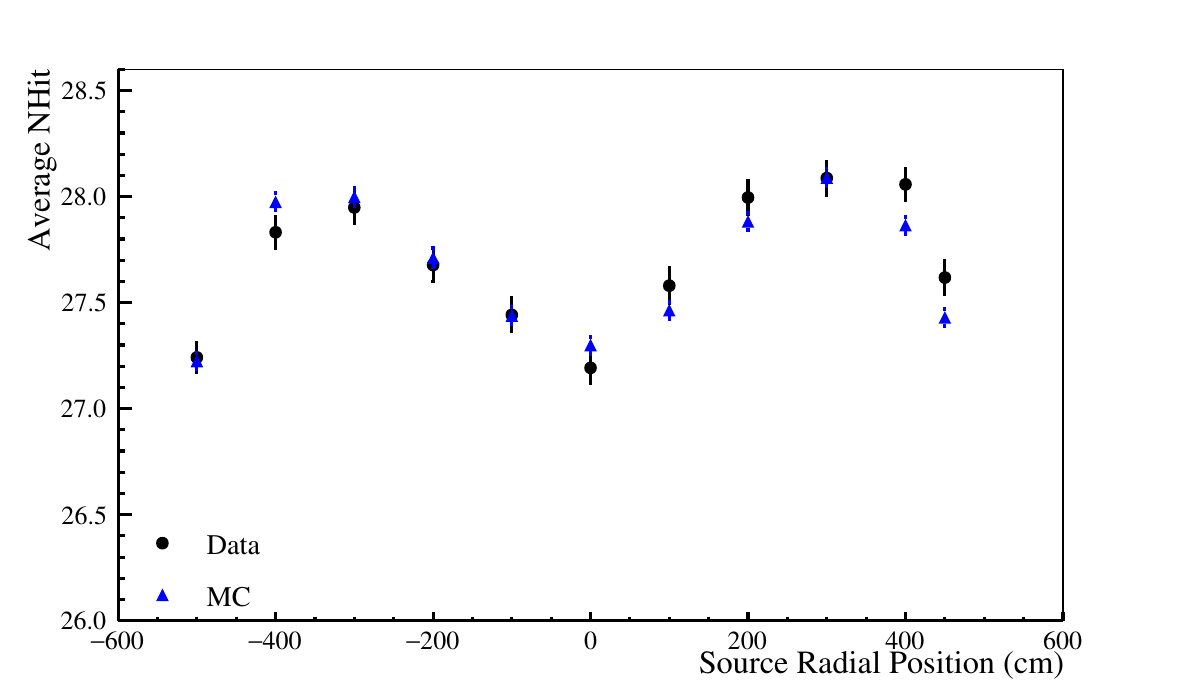}
	\includegraphics[width=.8\textwidth]{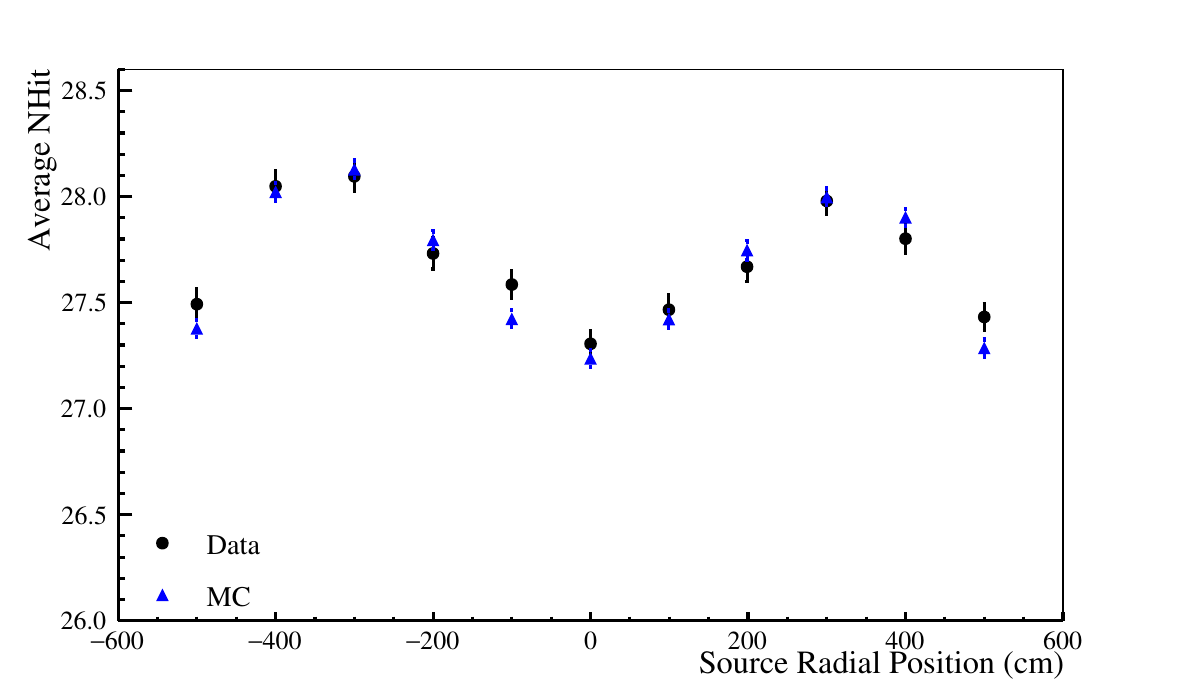}
	\\
	\includegraphics[width=.8\textwidth]{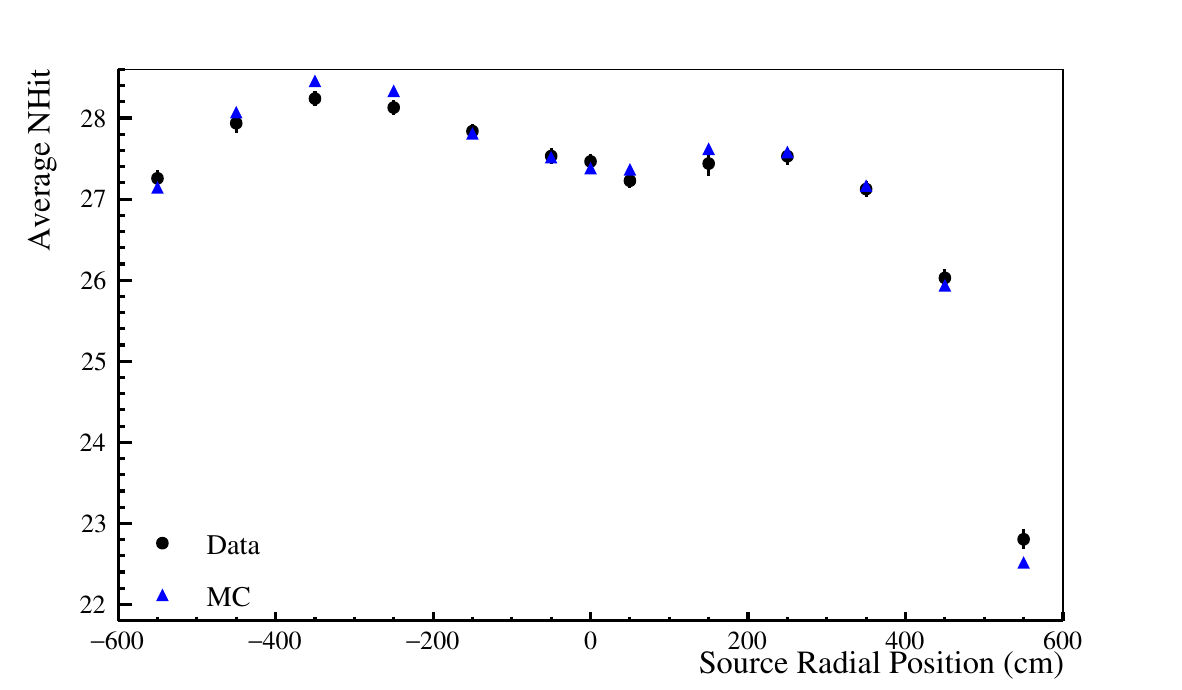}
	\caption{\label{fig:n16} Comparison between the mean number prompt of hits of the $^{16}$N source in data (black) and Monte Carlo (blue), as a function of axial position along the horizontal \emph{x} axis (top), the horizontal \emph{y} axis (middle) and the vertical \emph{z} axis (bottom).}
\end{figure}

Figure \ref{fig:n16} shows the comparison of the mean number of prompt hits in data and simulation as a function of the $^{16}$N source position along two horizontal axes and the vertical axis of the detector. As the calibration source moves away from the center, slightly more prompt light is collected relative to the central position. However, in positions closer to the AV, the average number of hits decreases. At high positions, this decrease is more accentuated due to the complex optics of the AV neck. 

\begin{figure}[htbp]
	\centering
	\includegraphics[width=.86\textwidth]{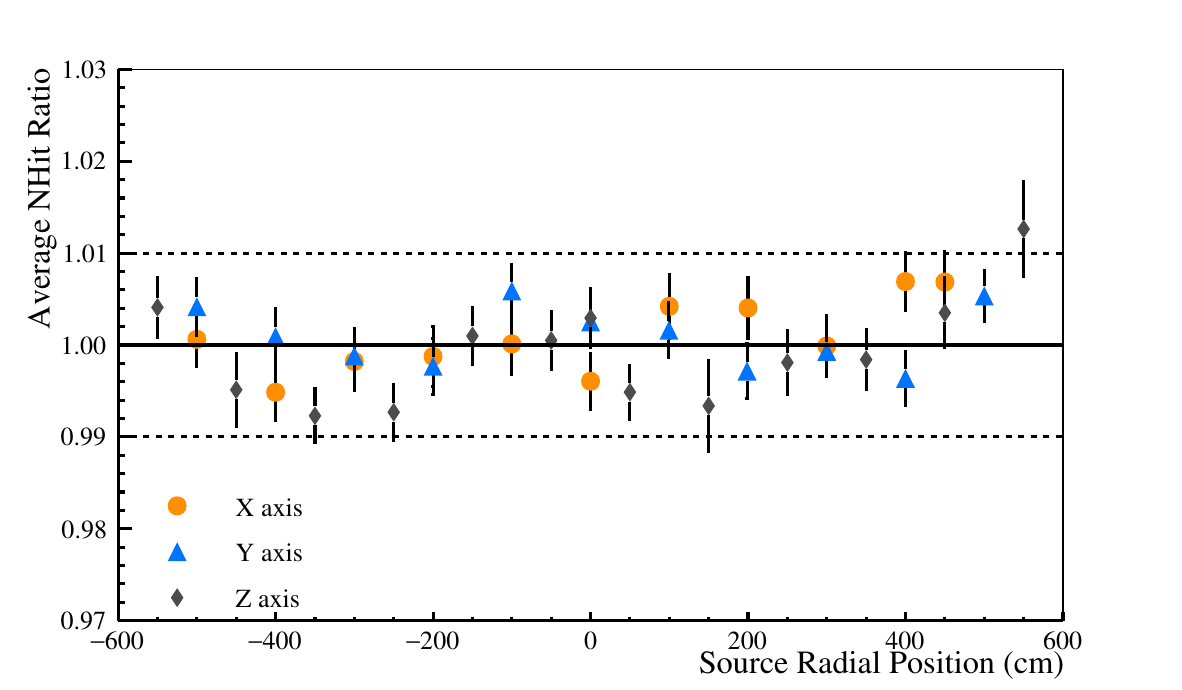}
	\caption{\label{fig:n16Ratios} Ratio of the mean prompt number of hits of the $^{16}$N source in data over Monte Carlo, as a function of axial position along the horizontal \emph{x} and \emph{y} axes and the vertical \emph{z} axis, in orange, blue and grey, correspondingly. The horizontal dashed lines denote 1\% deviations.}
\end{figure}

\begin{figure}[htbp]
	\centering
	\includegraphics[width=.75\textwidth]{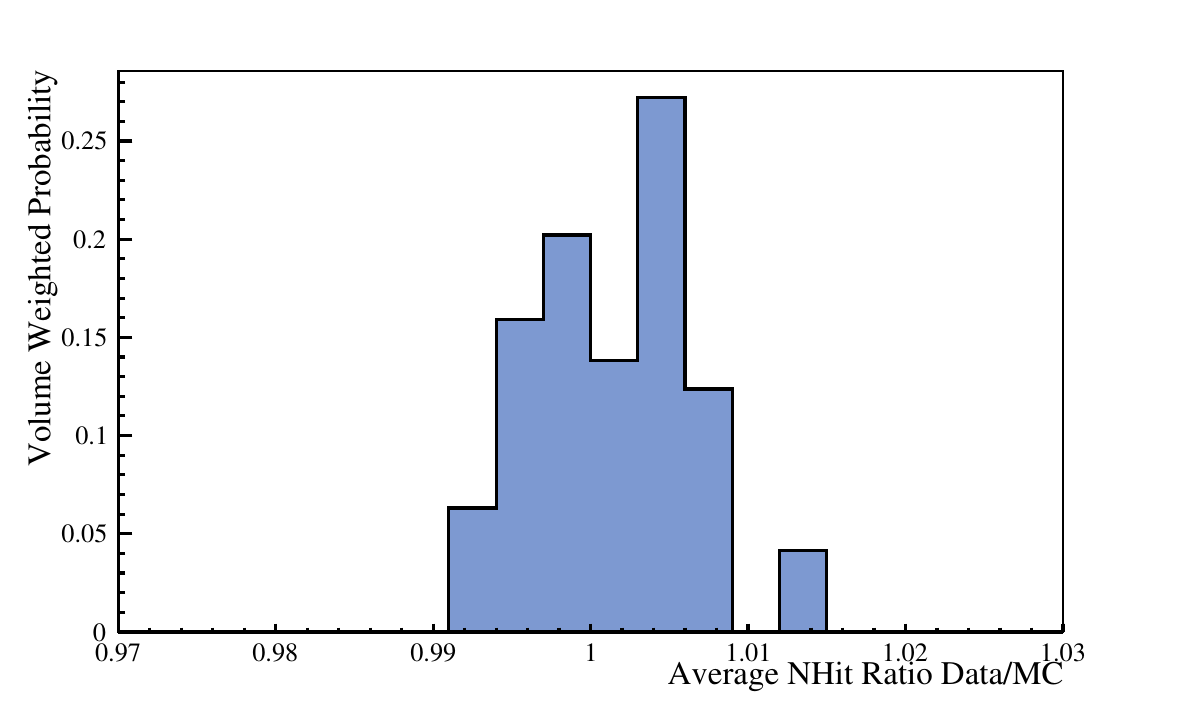}
	\caption{\label{fig:volWeightRatioDist} Distribution of the $^{16}$N prompt $N_\text{hit}$ ratios between data and Monte Carlo weighted by volume, up to a source deployment radius of 5.5 m.}
\end{figure}

Figure \ref{fig:n16Ratios} shows the ratio between the prompt hits in data and simulation as a function of the $^{16}$N source axial position. An agreement better than 1\% is found, validating the measured optical parameters. It is worth noting the good agreement in the $+z$-axis, in particular at larger axial positions where the data are affected by the optical properties of the acrylic vessel neck, which is not as UV transparent as the rest of the AV. Figure \ref{fig:volWeightRatioDist} shows the volume weighted distribution of the ratios, up to a radius of 5.5 m, from which it is possible to evaluate the contribution of the $N_\text{hit}$ position dependence to the energy scale systematic uncertainty. Adding the mean offset and the distribution width yields an uncertainty for the position dependence of the $N_\text{hit}$ of 0.6\%.

\section{Conclusion}
\label{section:conclusions}

The laserball source was successfully deployed during the SNO+ water phase, and all data acquired during two main calibration campaigns were analyzed, allowing a detailed and precise characterization of the optical effects of the detector media and PMTs at different wavelengths. The water phase of SNO+ provided a unique opportunity to obtain precise measurement of the media attenuations. The internal and external AV regions, both filled with ultra-pure water, were treated as the same material, which allowed to break the correlation between the external water and acrylic attenuations, and allowing an in-situ measurement of the latter. Including the external laserball data in the analysis contributed further to the sensitivity of this analysis, by allowing to scan the optical properties of the PMT-concentrator assembly in a wide range of light incidence angles.

The data from internal and external laserball scans were analyzed together, allowing an \textit{in situ} measurement of the effective acrylic attenuation and the angular response of the PMT-concentrator assembly at incidence angles above 45 degrees for the first time. Additionally, the attenuation coefficients of water were measured. These measurements were propagated to the detector simulation model, and comparisons between the $^{16}$N tagged gamma source data and Monte Carlo showed a good agreement, yielding an uncertainty for the position dependence of the energy scale of 0.6\% for the $N_\text{hit}$ as energy estimator, across all internal positions scanned by the $^{16}$N source. 

\newpage


\acknowledgments

Capital construction funds for the SNO+ experiment were provided by the Canada Foundation for Innovation (CFI) and matching partners. This research was supported by: Canada: Natural Sciences and Engineering Research Council, the Canadian Institute for Advanced Research (CIFAR), Queen’s University at Kingston, Ontario Ministry of Research, Innovation and Science, Alberta Science and Research Investments Program, National Research Council, Federal Economic Development Initiative for Northern Ontario (FedNor), Northern Ontario Heritage Fund Corporation, Ontario Early Researcher Awards; US: Department of Energy Office of Nuclear Physics, National Science Foundation, the University of California, Berkeley, Department of Energy National Nuclear Security Administration through the Nuclear Science and Security Consortium; UK: Science and Technology Facilities Council (STFC), the European Union’s Seventh Framework Programme under the European Research Council (ERC) grant agreement, the Marie Curie grant agreement; Portugal: Funda\c{c}\~{a}o para a Ci\^{e}ncia e a Tecnologia (FCT-Portugal); Germany: the Deutsche Forschungsgemeinschaft; Mexico: DGAPA-UNAM and Consejo Nacional de Ciencia y Tecnolog\'{i}a.

We would like to thank SNOLAB and its staff for support through underground space, logistical and technical services. SNOLAB operations are supported by CFI and the Province of Ontario Ministry of Research and Innovation, with underground access provided by Vale at the Creighton mine site.

This research was enabled in part by support provided by WestGRID (www.westgrid.ca) and ComputeCanada (www.computecanada.ca), in particular computer systems and support from the University of Alberta (www.ualberta.ca) and from Simon Fraser University (www.sfu.ca); and by the GridPP Collaboration, in particular computer systems and support from Rutherford Appleton Laboratory\cite{gridpp1,gridpp2}. Additional high-performance computing was provided through the “Illume” cluster funded by CFI and Alberta Economic Development and Trade (EDT) and operated by ComputeCanada and the Savio computational cluster resource provided by the Berkeley Research Computing program at the University of California, Berkeley (supported by the UC Berkeley Chancellor, Vice Chancellor for Research, and Chief Information Officer). Additional long-term storage was provided by the Fermilab Scientific Computing Division. Fermilab is managed by Fermi Research Alliance, LLC (FRA) under Contract with the U.S. Department of Energy, Office of Science, Office of High Energy Physics.

\end{document}